\documentclass[11pt, english]{article}

\usepackage{xparse}
\usepackage{xspace}

\usepackage[T1]{fontenc}
\usepackage{lmodern}
\usepackage{alphabeta}
\DeclareUnicodeCharacter{1EC5}{\~{ê}}
\usepackage{stmaryrd}
\usepackage{babel}
\usepackage{csquotes}

\usepackage{mathtools}
\usepackage{amsfonts,amssymb}
\usepackage{ebproof}
	\NewDocumentCommand \drule {m} {\ensuremath{\mathsf{(#1)}}}
	\ebproofset{
	right label template=\small\drule{\inserttext},
	}

\usepackage{geometry}
\usepackage{amsthm,thmtools}
\AtEndPreamble{
\newtheorem{theorem}{Theorem}
\newtheorem{lemma}[theorem]{Lemma}
\newtheorem{corollary}[theorem]{Corollary}

\theoremstyle{definition}
\newtheorem{definition}[theorem]{Definition}
\theoremstyle{remark}
\newtheorem{remark}[theorem]{Remark}
\newtheorem{example}[theorem]{Example}
}

\usepackage{todonotes}
	\setuptodonotes{size=\footnotesize}

\usepackage[style=alphabetic]{biblatex}
\usepackage{hyperref}
\hypersetup{
	colorlinks,
	allcolors=blue,
	pdftitle={Graded Intersection Types},
	pdfauthor={Rémy Cerda, Ugo Dal Lago}
}
\usepackage[capitalise]{cleveref}
\usepackage{./thmitemref}

\NewDocumentCommand \eg 			{} 		{\emph{e.g.}\@\xspace}
\NewDocumentCommand \ie 			{} 		{\emph{i.e.}\@\xspace}
\NewDocumentCommand \ifandonlyif 	{} 		{iff\xspace}
\NewDocumentCommand \resp 			{} 		{resp.\@\xspace}
\NewDocumentCommand \wrt 			{} 		{wrt.\@\xspace}

\NewDocumentCommand \eqdef 			{} 		{\coloneqq}

\NewDocumentCommand \set 			{smo} 	
	{\IfBooleanT{#1}{\left}\{ #2
	\IfValueT{#3}{\;\IfBooleanT{#1}{\middle}|\; #3}
	\IfBooleanT{#1}{\right}\}}
\RenewDocumentCommand \emptyset		{}		{\varnothing}
\NewDocumentCommand \subsetf		{}		{\subseteq_{\mathrm f}}
\NewDocumentCommand \cardinal		{m}		{\left|#1\right|}
\NewDocumentCommand \natset			{}		{\mathbb N}

\DeclareMathOperator \support				{supp}

\NewDocumentCommand \Semiring 		{s} 	{\mathbf S\IfBooleanT{#1}{^*}}
\NewDocumentCommand \Bool	 		{s} 	{\mathbf 2\IfBooleanT{#1}{^*}}
\NewDocumentCommand \Nat	 		{s} 	{\mathbf N\IfBooleanT{#1}{^*}}
\NewDocumentCommand \Tropical 		{s} 	{\mathbf T\IfBooleanT{#1}{^*}}

\NewDocumentCommand \+ 				{o} 	{+\IfValueT{#1}{_{#1}}}
\NewDocumentCommand \SUM 			{} 		{\sum}
\NewDocumentCommand \0 				{o} 	{0\IfValueT{#1}{_{#1}}}
\RenewDocumentCommand \* 			{o} 	{\times\IfValueT{#1}{_{#1}}}
\NewDocumentCommand \1 				{o} 	{1\IfValueT{#1}{_{#1}}}
\AtBeginDocument{
\RenewDocumentCommand \<			{t=o}
	{\IfBooleanTF{#1}{\le}{<} \IfValueT{#2}{_{#2}}}
\RenewDocumentCommand \>			{t=o}
	{\IfBooleanTF{#1}{\ge}{>} \IfValueT{#2}{_{#2}}}
}

\NewDocumentCommand \msets 			{O{\Nat}ms}
	{\mathcal M_{#1}(#2) \IfBooleanT{#3}{^*}}
\NewDocumentCommand \fmsets 		{O{\Nat}ms}
	{\mathcal M^{\mathrm f}_{#1}(#2) \IfBooleanT{#3}{^*}}
\NewDocumentCommand \lfmsets 		{O{\Nat}ms}
	{\mathcal M^{\mathrm{lf}}_{#1}(#2) \IfBooleanT{#3}{^*}}

\NewDocumentCommand \ms 			{m} 	{\mathsf{#1}}
\NewDocumentCommand \gr 			{mm} 	{#1:#2}

\NewDocumentCommand \msplus 		{} 		{\oplus}
\NewDocumentCommand \mssum 			{} 		{\bigoplus}
\NewDocumentCommand \mstimes 		{} 		{\odot}

\NewDocumentCommand \Var 			{} 		{\mathcal V}
\NewDocumentCommand \lamterms 		{} 		{\mathbf\Lambda}

\DeclareMathOperator \fv 					{fv}
\NewDocumentCommand \subst 			{O{x}m} {[#2/#1]}

\NewDocumentCommand \bred 			{s} {\mathrel{\rightarrow_\beta}\IfBooleanT{#1}{^*}}

\NewDocumentCommand \WN 			{} 		{\mathcal{WN}}
\NewDocumentCommand \HN 			{} 		{\mathcal{HN}}
\NewDocumentCommand \HHN 			{} 		{\mathcal{HN}^{+}}

\NewDocumentCommand \Atoms 			{} 		{\mathcal A}
\NewDocumentCommand \lto 			{} 		{\multimap}
\NewDocumentCommand \inlater 		{m} 	{\in_{#1}}

\NewDocumentCommand \itypes 		{t!O{\Semiring}} 
	{\IfBooleanT{#1}{\mathord{!}}\mathsf I_{\smash{#2}}}
\NewDocumentCommand \ritypes 		{t!O{\Semiring}} 
	{\IfBooleanT{#1}{\mathord{!}}\mathsf{rI}_{\smash{#2}}}
\NewDocumentCommand \lfitypes 		{t!O{\Semiring}} 
	{\IfBooleanT{#1}{\mathord{!}}\mathsf I^{\mathrm{lf}}_{\smash{#2}}}
\NewDocumentCommand \lfritypes 		{t!O{\Semiring}} 
	{\IfBooleanT{#1}{\mathord{!}}\mathsf{rI}^{\mathrm{lf}}_{\smash{#2}}}

\NewDocumentCommand \derivD 		{} 		{\mathcal D}
\NewDocumentCommand \derives 		{} 		{\blacktriangleright}

\NewDocumentCommand \redinterp 		{sO{\rho}m}
	{\IfBooleanT{#1}{\left}\llbracket #3 \IfBooleanT{#1}{\right}\rrbracket_{#2}}
\NewDocumentCommand \rcstyle		{m}		{\mathcal{#1}}

\NewDocumentCommand \fmsgrades		{O{\Semiring}} {[#1]}
\NewDocumentCommand \lfmsgrades		{O{\Semiring}} {[[#1]]}
\NewDocumentCommand \Tropicalhat 	{}		{\lfmsgrades[\Tropical]}
\NewDocumentCommand \mscurr			{m}		{\tilde{\ms{#1}}}
\NewDocumentCommand \msgset			{m}		{[[#1]]}
\NewDocumentCommand \collapse		{}		{\kappa}

\NewDocumentCommand \Monoid			{}		{\mathbf M}
\NewDocumentCommand \Quantities 	{} 		{\mathbf Q}

\NewDocumentCommand \dist			{so}
	{\mathbf{d} \IfBooleanT{#1}{^!} \IfValueT{#2}{_{#2}}}

\NewDocumentCommand \trunc 			{O{d}m}	{\lfloor #2\rfloor_{#1}}

\begin{document}

\title{Graded Intersection Types}
\author{%
	Rémy Cerda%
		\footnote{DISI, Università di Bologna, Bologna, Italy;
		Centre INRIA d'Université Côte d'Azur, Sophia-Antipolis, France.}
		\footnote{Contact email: \href{mailto:Remy.Cerda@math.cnrs.fr}{Remy.Cerda@math.cnrs.fr}.}
	\and
	Ugo Dal Lago\footnotemark[1]
}
\date{}
\maketitle

\paragraph{Intersection Types.}
\emph{Intersection types} \autocite{Coppo.Dez.78} were introduced as a refinement of simple types for the λ-calculus where arrow types $\sigma \to \tau$ are replaced with \[\set{\sigma_1,\dots,\sigma_n} \to \tau,\] \ie the domain of the arrow becomes \emph{a finite set of types} (presented as an intersection $\sigma_1 \cap \ldots \cap \sigma_n$ in the original formalism, hence the terminology).
The key feature of intersection type systems is that they enjoy not only subject reduction, as any reasonable type system, but also \emph{subject expansion}: if $M \bred N$, any typing of~$N$ can be turned backwards into a typing of~$M$.
As a consequence, intersection types have given rise to neat semantic counterparts (in particular as filter models \autocite{Barendregt.Cop.Dez.83}) and, on the operational side, to characterisations of computational properties: while simple types only \emph{guarantee} termination, intersection types allow to \emph{characterise}, \eg, weak \autocite{Coppo.Dez.78,Krivine.90} or head \autocite{Salle.80,Salle.80a} normalisation.

\emph{Non-idempotent} intersection types \autocite{Gardner.94,Kfoury.00} renewed this approach by relaxing indempotence of intersection, \ie $\sigma \cap \sigma = \sigma$ no longer holds; equivalently arrow types are presented with a finite \emph{multiset} of types as a domain, as in \[[\sigma_1,\dots,\sigma_n] \to \tau.\]
This shift unveiled a strong connexion with various program analysis tools related to linear logic: in particular, non-idempotent typings are in close correspondence with the standard relational model of the λ-calculus \autocite{Bucciarelli.Ehr.Man.07,deCarvalho.07,deCarvalho.17}, and with the multilinear approximation of λ-terms given by Taylor expansion \autocite{Ehrhard.Reg.08,Tsukada.Ong.16}.
As a consequence, the characterisations of operational properties (that are made much simpler to prove in a non-idempotent setting, thanks to linearity) could be turned into resource-sensitive results providing quantitative information and exact bounds on, \eg, resource usage or execution time \autocite{deCarvalho.17}.

\paragraph{Graded Types.}
In the same years, another thread of the literature studied a different enrichment of the domain of arrow types allowing to control the usage of resources by functional programs. Grounded in linear logic again \autocite{Girard.87} and in its decomposition of the arrow $\sigma \to \tau$ into $!\sigma \lto \tau$ (where the exponential modality $!$ allows for erasure or duplication of its content while the linear arrow $\lto$ requires a single use of the argument), bounded linear logic \autocite{Girard.Sce.Sco.92} introduced annotated exponentials, as in \[!_a \sigma \lto \tau,\] where $a$ is intuitively a quantity indicating \emph{how many times} the argument of a program may be used. \emph{Graded types} \autocite{Brunel.Gab.Maz.Zda.14,Ghica.Smi.14} then simplified and generalised this idea by allowing the annotation $a$ to be taken in any semiring, providing control on various aspects of the input of programs; this gave rise to a rich literature on \emph{graded coeffects}. 

\paragraph{Contribution.}
The present work aims at reconciling these two threads along the lines of a simple observation: the domains of the arrow in idempotent and non-idempotent intersection types (\ie, sets and multisets of types) can be seen as maps from types to booleans and from types to natural numbers, respectively, \ie as maps associating grades to types. We introduce \emph{graded intersection types}, a generalisation of intersection types where arrow types can be written as \[ [\gr{\sigma_1}{a_1}, \dots, \gr{\sigma_n}{a_n}] \lto \tau. \] The standard idempotent and non-idempotent settings are the instances where the grades $a_i$ are taken to be booleans or natural integers, respectively.

In \cref{sec:finite} we first discuss some algebraic properties of the grading semiring~$\Semiring$, and we introduce what we call $\Semiring$-multisets; then we define the graded intersection type system $\itypes$ and its relevant fragment.
In \cref{sec:sr-se} we establish that, under suitable algebraic conditions on~$\Semiring$, the system enjoys subject reduction and subject expansion.
In \cref{sec:lf} we extend all the previous work to infinitary graded intersection types relaxing finite support of multisets of types (we consider \emph{locally finite} support instead) as well as inductive type formation (we consider guarded coinductive formation instead), giving rise to \emph{locally finite graded intersection types}; their instance with grades taken in the tropical semiring has been thoroughly studied in another work \autocite{Cerda.Dal.26}.
In \cref{sec:multiset-constr} we introduce the graded counterpart to switching from idempotent to non-idempotent intersection types: we replace single grades with \emph{multisets of grades}. This can actually be done within the formalism, since these multisets themselves form a semiring with all desired properties. This lifting offers all the usual benefits of non-idempotency and linearity (which is mostly left for future work in the current version of this note).
Finally in \cref{sec:operational} we provide or recall some operational characterisations allowed by graded intersection types. In particular, as expected, unrestricted finite typing in $\itypes$ characterises head normalisation.

\paragraph{Related Work.}
Previous investigations already provided a partial semantic counterpart to this work. In particular \textcite{Carraro.Ehr.Sal.10} showed that the graded exponential comonad associated to a resource semiring gives rise to a relational model of the λ-calculus (the standard relational model being built for the semiring of natural integers). \Textcite{Breuvart.Pag.15} showed how the models generated this way allow to interpret graded types, and \textcite{Breuvart.Ker.Mir.26} extended this approach to graded types for the differential λ-calculus. Although it is not explored yet, there should be a direct connection between these semantics and the syntax we introduce in this paper. Extending these semantic approaches to \emph{weak} resource semirings, \eg finding a relational model for boolean or tropical grades, is the subject of ongoing work. Extending our syntax with differentiation, or designing a graded Taylor approximation, is another promising perspective.

The infinitary intersection types that we introduce are reminiscent of the ones by \textcite{Vial.17,Vial.21}, and actually gives rise to a characterisation of the same property of hereditary head normalisation \autocite{Cerda.Dal.26}. As opposed to the rather involved external machinery in Vial's work, our guarded coinductive definition already provides all needed ingredients for the proof.

It is outside the scope of this note to recall the wide literature on graded type systems controlling the coeffects of functional programs in order to ensure good behaviours, or to bound quantitative measures of these behaviours. Let us only mention that one long-term goal of the present work is to build on this literature so as to provide \emph{characterisations} or \emph{exact bounds} by putting into work the good properties of intersection types.

\section{Finite Graded Intersection Type Systems}
\label{sec:finite}

\subsection{Algebraic Preliminaries}
\label{sec:finite:algebraic}

\begin{definition}[Semiring]
\label{def:semiring}
	A \emph{semiring} is a tuple $(\Semiring,\mathord{\+},\0,\mathord{\*},\1)$, where
	$(\Semiring,\mathord{\+},\0)$ is a commutative monoid,
	$(\Semiring,\mathord{\*},\1)$ is a monoid,
	multiplication distributes over addition on both sides, and
	$\0$ is absorbing for multiplication.
	Throughout this note we also assume that $\0\neq\1$.
	We define the algebraic preorder $\<=$ on $\Semiring$ by writing
	$a \<= b$ whenever $\exists c,\ b = a \+ c$.
	
\end{definition}

Standard examples include the natural, boolean and tropical semirings,
respectively defined by:
\[
	\Nat \eqdef (\natset, +, 0, \times, 1), \qquad
	\Bool \eqdef (\set{0,1}, \vee, 0, \wedge, 1), \qquad
	\Tropical \eqdef (\natset \cup \set{\infty}, \min, \infty, +, 0).
\]
where $\natset$ denotes the \emph{set} of natural integers.

For a semiring to give rise to a well-behaved intersection type system enjoying, in particular, subject reduction (see \cref{sec:sr-se}), we need to enforce some algebraic conditions. 

\begin{definition}[Resource Semiring]
	\label{def:resource-semiring}
	Given a semiring $(\Semiring,\mathord{\+},\0,\mathord{\*},\1)$, consider the conditions:
	\begin{flalign*}
		\text{\emph{Positivity}}\quad &
			a\+b=\0\implies a=b=\0, &\\
		\text{\emph{No Zero Divisors}}\quad &
			a\*b=\0\implies a=\0\text{ or }b=\0, \\
		\text{\emph{Discreteness}}\quad &
			a\+b=\1\implies a=\0\text{ or }b=\0, \\
		\text{\emph{Weak Discreteness}}\quad &
			a\+b=\1\implies a \>= 1 \text{ or } b \>= 1, \\
		\text{\emph{Additive Refinement}}\quad &
			a_1\+a_2=b_1\+b_2\implies \exists c_{11},c_{12},c_{21},c_{22}\in\Semiring, \\&
			\forall i\in\set{1,2},\ a_i=c_{i1}\+c_{i2},\qquad
			\forall j\in\set{1,2},\ b_j=c_{1j}\+c_{2j}, \\
		\text{\emph{Mixed Refinement}}\quad &
			a\*b=c_1\+c_2\implies \exists L\text{ finite},\ \exists(a_l)_{l\in L},\ 
			\exists(b_{il})_{i\in\set{1,2},l\in L}, \\&
			a=\SUM_{l\in L}a_l,\qquad
			\forall l\in L,\ b=b_{1l}\+b_{2l},\qquad
			\forall i\in\set{1,2},\ c_i=\SUM_{l\in L}a_l\*b_{il}.
	\end{flalign*}
	Then $\Semiring$ is said to be a \emph{(weak) resource semiring} whenever it is positive, has no zero divisors, is (weakly) discrete, additive refining and mixed refining. When contrasting the two notions, we also say \emph{strong resource semiring} for a resource semiring.
\end{definition}

The conditions for a resource semiring are already well-known in the literature:
they ensure that the $\Semiring$-semimodule monad distributes over the powerset monad \autocite{Clementino.Hof.Jan.13,Bonchi.San.22},
and they play a crucial role in the construction of an exponential comonad when designing graded versions of the relational and coherent models of linear logic \autocite{Carraro.Ehr.Sal.10,Breuvart.Pag.15,Moreau.Ngu.26}.

We depart from the terminology coined by \textcite{Carraro.Ehr.Sal.10} where resource semirings, additive refinement and mixed refinement were called \enquote{multiplicity semirings}, \enquote{additive splitting} and \enquote{multiplicative splitting}, respectively.
The first change aims at avoiding ambiguities with existing notions, and is borrowed from \textcite{Breuvart.Ker.Mir.26};
the latter two bring back the original and standard algebraic terminology \autocite{Tarski.49}.

The weak discreteness condition, as far as we know, is new; investigating how it can enter the semantic picture is ongoing work.

\begin{lemma}[$n$-ary Properties]
	\label{lem:nary}
	Let $\Semiring$ be a semiring that is positive and has no zero divisors. The following properties hold, with all the considered index sets being finite. Then:
	\begin{enumerate}
	\item \thmitemlabel{lem:nary:nary-discrete}
		If $\Semiring$ has discreteness, 
		then it has \emph{$n$-ary discreteness}:
		if $\SUM_{i \in I} a_i = \1$, then exactly one $a_i$ is non-zero (and is equal to~$\1$). \\
	If $\Semiring$ has weak discreteness and additive refinement,
		then it has \emph{$n$-ary weak discreteness}:
		if~$\SUM_{i\in I} a_i \>= \1$, then $\exists i,\ a_i \>= \1$.
	\item \thmitemlabel{lem:nary:nary-additive-refinement}
		If $\Semiring$ has additive refinement,
		then it has \emph{$n$-ary additive refinement}:
		\begin{gather*}
			\SUM_{i \in I} a_i=\SUM_{j \in J} b_j \implies
			\exists (c_{ij})_{i\in I,j\in J}, \quad
			\forall i \in I,\ a_i=\SUM_{j \in J} c_{ij}, \quad
			\forall j \in J,\ b_j=\SUM_{i \in I} c_{ij}.
		\end{gather*}
	\item \thmitemlabel{lem:nary:nary-mixed-refinement}
		If $\Semiring$ has additive and mixed refinement, 
		then it has \emph{$n$-ary mixed refinement}:
		\begin{multline*}
			a\*b=\SUM_{i \in I} c_i \implies \exists L,\ 
			\exists (a_l)_{l \in L},\ \exists (b_{il})_{i\in I,l\in L}, \\
			a=\SUM_{l\in L} a_l,\qquad
			\forall l\in L,\ b=\SUM_{i \in I} b_{il},\qquad
			\forall i \in I,\ c_i=\SUM_{l\in L} a_l\*b_{il},
		\end{multline*}
		as well as \emph{simultaneous $n$-ary mixed refinement}:
		\begin{multline*}
			\forall x\in X,\ a\*b_x=\SUM_{i\in I_x}c_{xi} \implies
			\exists L,\ 
			\exists (a_l)_{l \in L},\ 
			\exists (b_{xil})_{x\in X,i\in I_x,l\in L}, \\
			a=\SUM_{l\in L} a_l,\qquad
			\forall x\in X,\ \left(
				\forall l\in L,\ b_x=\SUM_{i \in I_x} b_{xil},\quad
				\forall i \in I_x,\ c_{xi}=\SUM_{l\in L} a_l\*b_{xil}
			\right).
		\end{multline*}
	\item  \thmitemlabel{lem:nary:nary-resource-semiring}
		$\Semiring$ is a (weak) resource semiring \ifandonlyif it has $n$-ary (weak) discreteness, $n$-ary additive refinement, and $n$-ary mixed refinement.
	\end{enumerate}
\end{lemma}

	\begin{proof}
	For \cref{lem:nary:nary-discrete,lem:nary:nary-additive-refinement} the proof is a straightforward induction on the size of the considered index sets, by writing $\SUM_i a_i = a_{i_0} \+ \SUM_{i \neq i_0} a_i$ and using the corresponding binary hypotheses on~$\Semiring$. The case where $I$ (or $J$) is the empty set is trivial for for \cref{lem:nary:nary-discrete} and given by positivity for \cref{lem:nary:nary-additive-refinement}.
	
	For \cref{lem:nary:nary-mixed-refinement}, we first prove $n$-ary mixed refinement, by induction. Write $a\*b = c_{i_0} \+ \sum_{i \neq i_0} c_i$. Mixed refinement and $n$-ary additive refinement give families of elements of $\Semiring$ such that:
	\begin{gather*}
		a = \SUM_l a_l,\qquad
		\forall l,\ b = b_{i_0l} \+ b'_{i_0l},\qquad
		c_{i_0} = \SUM_l a_l \* b_{i_0l}, \\
		\forall i\neq i_0,\ c_i=\SUM_l d_{il}, \qquad
		\forall l,\ a_l \* b'_{i_0,l} = \SUM_{i\neq i_0} d_{il}.
	\end{gather*}
	Apply induction on the last equality. The case where $I$ is the empty set is given by the absence of zero divisors.
	For simultaneous $n$-ary mixed refinement, apply $n$-ary mixed refinement separately for each $x$, then take a common additive refinement of the finitely many decompositions of $a$.
	
	For \cref{lem:nary:nary-resource-semiring}, there only remains to notice that positivity and absence of zero divisors correspond to the $0$-ary versions of additive and mixed refinement, respectively.
	\end{proof}

\begin{remark}
	\label{rem:binary-mixed-refinement}
	The finite set $L$ in (binary) mixed refinement cannot in general be restricted to two indices. For example, in $\Nat[u_1,u_2,u_3,v_1,v_2]$, put
	\[
		a=u_1+u_2+u_3,\qquad b=v_1+v_2,\qquad
		c_1=u_1v_1+u_2v_2,\qquad
		c_2=u_1v_2+u_2v_1+u_3v_1+u_3v_2.
	\]
	Suppose that these admit a mixed refinement. Since each $u_i$ has coefficient $1$ in $a$, it would belong to exactly one component $a_l$. The corresponding $b_{1l}$ must be $v_1$ if this component contains $u_1$, $v_2$ if it contains $u_2$, and $0$ if it contains $u_3$. These requirements are incompatible within a single component, so at least three components are necessary.
	
	Requiring $L=\set{1,2}$ is exactly condition (MS4) of \autocite{Carraro.Ehr.Sal.10} and is therefore strictly stronger than our mixed-refinement condition, which instead corresponds to (MS4') in the later \autocite{Breuvart.Pag.15}. For $n$-ary mixed refinement, the bound $\cardinal L\leq2^{\cardinal I-1}$ from \autocite[Lemma~2]{Carraro.Ehr.Sal.10} assumes the stronger binary condition (MS4), and does not hold under our assumptions.
\end{remark}

\begin{lemma}
	\label{lem:basic-resource}
	$\Nat$ is a resource semiring.
	$\Bool$ and $\Tropical$ are weak resource semirings.
\end{lemma}

For another example, the polynomial semiring $\Nat[X]$ is a strong resource semiring: apply \cref{lem:monoid} to the free monoid on one generator. The same argument applies to any set of commuting indeterminates, using the free commutative monoid.

\medskip

Finally, the usual notion of multisets can be generalised to functions having a semiring as their codomain, as follows:

\begin{definition}[$\Semiring$-Multisets]
\label{def:msets}
	For any set $X$, an \emph{$\Semiring$-multiset} on $X$ is a map $\ms m:X\to\Semiring$. It is \emph{finite} if its support $\support(\ms m)\eqdef\set{x\in X}[\ms m(x)\neq\0]$ is finite. We denote the set of finite $\Semiring$-multisets by $\fmsets[\Semiring]{X}$.
	Given finite families $(x_i)_{i\in I}$ in $X$ and $(a_i)_{i\in I}$ in $\Semiring$, we write $[\gr{x_i}{a_i}]_{i\in I}$ for the $\Semiring$-multiset $\ms m:x\mapsto\SUM_{i\in I,x_i=x}a_i$. In the following we always try to keep the $x_i$ pairwise distinct, so that the definition boils down to $\ms m(x_i)\eqdef a_i$, and $\ms m(x)\eqdef\0$ for $x\notin\set{x_i}[i\in I]$. When $I=\set{1,\dots,n}$, we also write $[\gr{x_1}{a_1},\dots,\gr{x_n}{a_n}]$. In particular, $[]$ is the \emph{empty $\Semiring$-multiset}.
\end{definition}

The operations of $\Semiring$ induce an internal sum and an external product on $\fmsets[\Semiring]{X}$:
\[
	(\ms m\+\ms n)(x)\eqdef\ms m(x)\+\ms n(x),\qquad
	(a\*\ms m)(x)\eqdef a\*\ms m(x).
\]
These operations preserve finite support. For $\Nat$ and $\Bool$, $\fmsets[\Nat]{X}$ and $\fmsets[\Bool]{X}$ are the sets of finite multisets and finite subsets of $X$, respectively.

\subsection{The Typing Systems}
\label{sec:finite:typing}

Let $\lamterms$ be the set of pure λ-terms, taken up to $\alpha$-conversion, built from a countable set $\Var$ of variables. We write $M\subst N$ for capture-avoiding substitution, $M\bred N$ for one β-reduction step, and $M\bred* N$ for a finite β-reduction sequence.

We now introduce a variant of intersection type systems where intersections of types, instead of being sets of types (idempotent intersections) or multisets of types (non-idempotent intersections), are finite $\Semiring$-multisets of types.

\begin{definition}[$\Semiring$-Intersection Types]
	\label{def:itypes}
	Let $\Atoms$ be a countable set of atoms. The sets $\itypes[\Semiring]$ of (finite) \emph{$\Semiring$-intersection types} and $\itypes![\Semiring]\subset\fmsets[\Semiring]{\itypes[\Semiring]}$ of \emph{$\Semiring$-intersection multi-types} are defined by mutual induction by the following formation rules:
	\begin{gather*}
	\begin{prooftree}
		\hypo{\alpha\in\Atoms}
		\infer1[\Atoms]{\alpha\in\itypes[\Semiring]}
	\end{prooftree}
	\qquad
	\begin{prooftree}
		\hypo{\ms m\in\itypes![\Semiring]}
		\hypo{\tau\in\itypes[\Semiring]}
		\infer2[\lto]{\ms m\lto\tau\in\itypes[\Semiring]}
	\end{prooftree}
	\qquad
	\begin{prooftree}[center]
		\hypo{[\gr{(\sigma_i\in\itypes[\Semiring])}{a_i}]_{i\in I}}
		\infer1[!]{[\gr{\sigma_i}{a_i}]_{i\in I}\in\itypes![\Semiring]}
	\end{prooftree}
	\end{gather*}
	where $I$ is finite. Notice that the formation rule \drule{!} has a finite $\Semiring$-multiset of hypotheses.
\end{definition}

Typing judgments are of the shape $\Gamma \vdash M:\sigma$, with auxiliary judgments of the shape $\Gamma \vdash M:\ms m$. The context $\Gamma$ is a map $\Var \to \itypes![\Semiring]$ whose support $\set{x \in \Var}[\Gamma(x) \neq []]$ is a finite set $\set{x_1,\dots,x_n}$. We will present $\Gamma$ in the usual way, as a list $x_1:\ms{m}_1, \dots, x_n:\ms{m}_n$ with pairwise distinct $x_i$. The operations induced on $\fmsets[\Semiring]{\itypes[\Semiring]}$ by the operations of $\Semiring$ are again mapped on contexts:
\[ (\Gamma \+ \Gamma') : x \mapsto \Gamma(x) \+ \Gamma'(x) \qquad
	(a \* \Gamma) : x \mapsto a \* \Gamma(x). \]

\begin{definition}[Typing] \label{def:itypes-typing}
	A λ-term $M$ is \emph{typable in $\itypes[\Semiring]$} whenever there is a derivation $\derivD$ with conclusion $\Gamma \vdash M : \sigma$ in the following (inductive) system of typing rules:
	\begin{gather*}
		\begin{prooftree}
		\infer0[ax]{ \Gamma \+ x:[\gr{\sigma}{\1}] \vdash x:\sigma }
		\end{prooftree}
		\qquad\qquad
		\begin{prooftree}
		\hypo{ \Gamma, x:\ms{m} \vdash M:\tau }
		\infer1[\lto_i]{ \Gamma \vdash λx.M : \ms{m} \lto \tau }
		\end{prooftree}
		\\[\topsep]
		\begin{prooftree}
		\hypo{ \Gamma \vdash M : \ms{m} \lto \tau }
		\hypo{ \Gamma' \vdash N : \ms{m} }
		\infer2[\lto_e]{ \Gamma \+ \Gamma' \vdash MN : \tau }
		\end{prooftree}
		\qquad
		\begin{prooftree}
		\hypo{ \left[ \gr {(\Gamma_i \vdash N:\sigma_i)} {a_i} \right]
			_{i \in I} }
		\infer1[!]{ \SUM_{i \in I} a_i \* \Gamma_i
			\vdash N :
			[\gr{\sigma_i}{a_i}]_{i \in I}
		}
		\end{prooftree}
	\end{gather*}
	where $I$ is finite, for some context $\Gamma$ and $\Semiring$-intersection type $\sigma$.
	We then write $\derivD \derives \Gamma \vdash M : \sigma$, and simply $\derives \Gamma \vdash M : \sigma$ to say that such a $\derivD$ exists.
\end{definition}

\begin{definition}[Relevant Typing]
\label{def:relevant}
	A λ-term $M$ is typable in the \emph{relevant system} $\ritypes$ whenever there is $\derivD \derives \Gamma \vdash M : \sigma$ in the system where \drule{ax} is replaced with the following typing rule:
	\[
		\begin{prooftree}
		\infer0[rax]{x:[\gr\sigma\1]\vdash x:\sigma}.
		\end{prooftree}
	\]
	We use $\vdash$ and $\derives$ for both systems, specifying the system when necessary.
\end{definition}

Thus $\itypes$ permits unused resources, including additional types on the variable in an axiom; $\ritypes$ does not.
For example, the original idempotent system from \textcite{Coppo.Dez.78} is  $\itypes[\Bool]$, 
while the non-idempotent system presented in the survey by \textcite{Bucciarelli.Kes.Ven.17} is $\ritypes[\Nat]$.
Both systems derive, \eg,
\[
	\vdash\lambda x.x(xx):
	[\gr{[\gr\alpha\1]\lto\alpha}{\1\+\1},\gr\alpha\1]\lto\alpha.
\]
Over $\Nat$ the coefficient $\1\+\1$ counts two uses; over $\Bool$ these uses are identified. In the tropical instance of the above typing, both $1$ and $1+1$ become $0$.

\begin{lemma}[Weakening and Relevance]
\label{lem:weakening}
	The following properties hold:
	\begin{enumerate}
	\item \thmitemlabel{lem:weakening:weakening}
		In $\itypes$, if $\derives \Gamma\vdash M:\sigma$, then $\derives \Gamma\+\Delta\vdash M:\sigma$.
	\item \thmitemlabel{lem:weakening:remove-useless-variables}
		In $\itypes$, if $\derives \Gamma, x:\ms m\vdash M:\sigma$ and $x\notin\fv(M)$, then $\derives \Gamma \vdash M:\sigma$.
	\item \thmitemlabel{lem:weakening:relevance}
		In $\ritypes$, if $\derives \Gamma\vdash M:\sigma$, then $\support(\Gamma)\subseteq\fv(M)$.
	\end{enumerate}
\end{lemma}

	\begin{proof}
	For \cref{lem:weakening:weakening}, propagate $\Delta$ through abstractions and through the left premise of applications, adding it to the axiom context. Choose bound variables outside $\support(\Delta)$, using α-equivalence. 
	\Cref{lem:weakening:remove-useless-variables,lem:weakening:relevance} are by induction on the derivation.
	\end{proof}

Notice that \cref{lem:weakening:weakening} concerns typing judgements only: arbitrary weakening of multi-typing judgments is not admissible. More precisely, the only contexts that can be added to a derivation whose conclusion is the typing rule \drule{!} are of the form $\sum_{i\in I} a_i \* \Delta_i$. In particular a derivation $\derives \mbox{} \vdash M:[]$ cannot be weakened (unless $\Semiring$ lacks positivity).
One can easily show that Items~2 and~3, instead, also apply to multi-typing judgements.

\section{Subject Reduction and Expansion}
\label{sec:sr-se}

The key feature of usual intersection types is that they enjoy not only subject reduction, but also subject expansion. We now show that this is still the case for $\Semiring$-intersection types, relying on the algebraic conditions introduced in \cref{sec:finite:algebraic}. We first show how these conditions come into play in the proof theory of our typing systems (\cref{sec:sr-se:algebraic}). Then we prove subject reduction and expansion for~$\itypes$ (\cref{sec:sr-se:proof}). Finally we show how the proof restricts to the relevant system~$\ritypes$, using strong discreteness (\cref{sec:sr-se:relevant}).

\subsection{Refining Typing Derivations}
\label{sec:sr-se:algebraic}

We work in $\itypes$; in particular, all types, all families and all index sets considered below are finite.

The following three lemmas transport the three items of \cref{lem:nary} to (contexts of) multi-types derivations. We isolate these properties in order to highlight where the algebraic conditions will come into play in the subject reduction proof that will follow (in \cref{sec:sr-se:proof}).

\begin{lemma}[Discreteness on Derivations]
	\label{lem:derivations-discreteness}
	If $\Semiring$ is positive, weakly discrete, and has additive refinement, then in $\itypes$,
	$\derives\Delta\vdash N:[\gr{\sigma}{\1}]$ implies $\derives\Delta\vdash N:\sigma$.
	If $\Semiring$ is positive and discrete, the same holds in $\ritypes$.
	In addition, the converse implications hold without any hypothesis on~$\Semiring$.
\end{lemma}

	\begin{proof}
	Start from a derivation
	\[	\begin{prooftree}
		\hypo{ \left[\gr{(\Delta_i\vdash N:\sigma)}{a_i}\right]_{i\in I} }
		\infer1[!]{ \Delta\vdash N:[\gr{\sigma}{\1}] }
		\end{prooftree} \]
	for some finite set~$I$, where
	$\SUM_{i\in I}a_i=\1$ and $\Delta=\SUM_{i\in I}a_i\*\Delta_i$.
	\thmitemref{lem:nary:nary-discrete} gives $i_0\in I$ with $a_{i_0}\>=\1$. Write $a_{i_0}=\1\+c$. Then
	\[
		\Delta=\Delta_{i_0}\+c\*\Delta_{i_0}\+\SUM_{i\in I\setminus\set{i_0}}a_i\*\Delta_i,
	\]
	and \thmitemref{lem:weakening:weakening} gives $\derives\Delta\vdash N:\sigma$.
	If $\Semiring$ is (strongly) discrete, the same reasoning gives a unique nonzero coefficient $a_{i_0}=\1$, so $\Delta=\Delta_{i_0}$ and weakening is unnecessary.
	Conversely just apply the typing rule \drule{!}.
	\end{proof}

\begin{lemma}[Additive Refinement on Derivations]
	\label{lem:derivations-additive-refinement}
	Assume that $\Semiring$ is positive and has additive refinement. For every finite family $(\ms m_j)_{j\in J}$, in both $\itypes$ and $\ritypes$,
	\begin{gather*}
		\text{if}\quad \derives\Delta\vdash N:\SUM_{j\in J}\ms m_j, \quad
		\text{then }\quad \exists(\Delta_j)_{j\in J},\quad
		\Delta=\SUM_{j\in J}\Delta_j,\qquad
		\forall j\in J,\ \derives\Delta_j\vdash N:\ms m_j.
	\end{gather*}
	In addition, the converse implication holds without any hypothesis on~$\Semiring$.
\end{lemma}

	\begin{proof}
	For the left-to-right implication, the hypothesis is obtained by the typing rule \drule{!} with premises $\Theta_i\vdash N:\sigma_i$ and coefficients $a_i$, so that
	\[
		\forall\sigma\in\itypes,\ \SUM_{\substack{i\in I\\\sigma_i=\sigma}}a_i=\SUM_{j\in J}\ms m_j(\sigma),\qquad
		\Delta=\SUM_{i\in I}a_i\*\Theta_i.
	\]
	Apply $n$-ary additive refinement (\thmitemref{lem:nary:nary-additive-refinement}) to the first equality, for each $\sigma \in \itypes$. We obtain $(c_{ij})_{i\in I,j\in J}$ such that
	\[
		\forall i\in I,\ a_i=\SUM_{j\in J}c_{ij},\qquad
		\forall j\in J,\ \forall \sigma\in\itypes,\ \ms m_j(\sigma)=\SUM_{\substack{i\in I\\\sigma_i=\sigma}}c_{ij}.
	\]
	Put $\Delta_j\eqdef\SUM_{i\in I}c_{ij}\*\Theta_i$ and form, for every $j\in J$,
	\[
		\begin{prooftree}
		\hypo{[\gr{(\Theta_i\vdash N:\sigma_i)}{c_{ij}}]_{i\in I}}
		\infer1[!]{\Delta_j\vdash N:\ms m_j}.
		\end{prooftree}
	\]
	Distributivity gives $\SUM_{j\in J}\Delta_j=\SUM_{i\in I}(\SUM_{j\in J}c_{ij})\*\Theta_i=\Delta$.

	For the converse, each $\Delta_j\vdash N:\ms m_j$ is obtained by the typing rule~\drule{!} with premises $\Theta_{ji}\vdash N:\sigma_{ji}$ and coefficients $a_{ji}$, indexed by $i\in I_j$. Write:
	\[
		\begin{prooftree}[center=false]
		\hypo{[\gr{(\Theta_{ji}\vdash N:\sigma_{ji})}{a_{ji}}]_{j\in J,i\in I_j}}
		\infer1[!]{\SUM_{j\in J}\Delta_j\vdash N:\SUM_{j\in J}\ms m_j}.
		\end{prooftree}	
		\qedhere	
	\]
	\end{proof}

\begin{lemma}[Mixed Refinement on Derivations]
	\label{lem:derivations-mixed-refinement}
	Assume that $\Semiring$ is positive, has no zero divisors, and has additive and mixed refinement. For all finite families $(a_k)_{k\in K}$ and $(\ms m_k)_{k\in K}$, in both $\itypes$ and $\ritypes$,
	\begin{align*}
		\text{if}\quad & \derives\Delta\vdash N:\SUM_{k\in K}a_k\*\ms m_k, \\
		\text{then}\quad &
		\exists(L_k)_{k\in K}\text{ finite sets},\
		\exists(a_{kl})_{k\in K,l\in L_k},\
		\exists(\Delta_{kl})_{k\in K,l\in L_k},\\
		& \forall k\in K,\ a_k=\SUM_{l\in L_k}a_{kl},\qquad
		\Delta=\SUM_{\substack{k\in K\\l\in L_k}}a_{kl}\*\Delta_{kl},\qquad
		\forall k\in K,l\in L_k,\ \derives\Delta_{kl}\vdash N:\ms m_k.
	\end{align*}
	In addition, the converse implication holds without any hypothesis on~$\Semiring$.
\end{lemma}

	\begin{proof}
	For the left-to-right implication, the hypothesis is obtained by the typing rule \drule{!} with premises $\Delta_i\vdash N:\sigma_i$ and coefficients $c_i$. Put $X\eqdef\set{\sigma_i}[i\in I]\cup\bigcup_{k\in K}\support(\ms m_k)$. Then
	\[
		\forall\sigma\in X,\ \SUM_{k\in K}a_k\*\ms m_k(\sigma)=\SUM_{\substack{i\in I\\\sigma_i=\sigma}}c_i,\qquad
		\Delta=\SUM_{i\in I}c_i\*\Delta_i.
	\]
	Apply $n$-ary additive refinement (\thmitemref{lem:nary:nary-additive-refinement}) to the first equality, for all $\sigma \in X$. There are coefficients $e_{ki}$ such that
	\[
		\forall\sigma\in X,k\in K,\ a_k\*\ms m_k(\sigma)=\SUM_{\substack{i\in I\\\sigma_i=\sigma}}e_{ki},\qquad
		\forall i\in I,\ c_i=\SUM_{k\in K}e_{ki}.
	\]
	For each $k\in K$, apply simultaneous $n$-ary mixed refinement (\thmitemref{lem:nary:nary-mixed-refinement}) to the family of equations indexed by $\sigma\in X$, with fibres $I_\sigma\eqdef\set{i\in I}[\sigma_i=\sigma]$. Since these fibres partition $I$, we index the resulting right coefficients by $i$ alone, writing them $b_{ikl}$. There are finite sets $L_k$ and coefficients $a_{kl},b_{ikl}$ such that for all $k \in K$,
	\begin{gather*}
		a_k=\SUM_{l\in L_k}a_{kl},\qquad
		\forall\sigma\in X,\ \forall l\in L_k,\ \ms m_k(\sigma)=\SUM_{\substack{i\in I\\\sigma_i=\sigma}}b_{ikl},\qquad
		\forall i\in I,\ e_{ki}=\SUM_{l\in L_k}a_{kl}\*b_{ikl}.
	\end{gather*}
	Finally, put $\Delta_{kl}\eqdef\SUM_{i\in I}b_{ikl}\*\Delta_i$ and form
	\[
		\begin{prooftree}
		\hypo{[\gr{(\Delta_i\vdash N:\sigma_i)}{b_{ikl}}]_{i\in I}}
		\infer1[!]{\Delta_{kl}\vdash N:\ms m_k}.
		\end{prooftree}
	\]
	Distributivity gives
	\[
		\SUM_{k,l}a_{kl}\*\Delta_{kl}=\SUM_{i\in I}\left(\SUM_{k,l}a_{kl}\*b_{ikl}\right)\*\Delta_i=\SUM_{i\in I}c_i\*\Delta_i=\Delta.
	\]
	Conversely, each $\Delta_{kl}\vdash N:\ms m_k$ is obtained by the typing rule \drule{!} with premises $\Theta_{kli}\vdash N:\tau_{kli}$ and coefficients $b_{ikl}$, indexed by $i\in I_{kl}$. Flatten the promotions:
	\[
		\begin{prooftree}[center=false]
		\hypo{[\gr{(\Theta_{kli}\vdash N:\tau_{kli})}{a_{kl}\*b_{ikl}}]_{k\in K,l\in L_k,i\in I_{kl}}}
		\infer1[!]{\Delta\vdash N:\SUM_{k\in K}a_k\*\ms m_k}.
		\end{prooftree}	
		\qedhere	
	\]
	\end{proof}

\subsection{Subject Reduction and Expansion for $\itypes$}
\label{sec:sr-se:proof}

In this subsection we work in $\itypes$, for which we prove subject reduction and subject expansion, relying as usual on two substitution lemmas. While the latter holds under no additional condition, the former needs $\Semiring$ to be a weak resource semiring -- as made apparent in the results of the previous Section.

\begin{lemma}[Forward Substitution]
	\label{lem:forward-substitution}
	Assume that $\Semiring$ is a weak resource semiring.
	If $\derives\Gamma,x:\ms m\vdash M:\sigma$ and $\derives\Delta\vdash N:\ms m$, then $\derives\Gamma\+\Delta\vdash M\subst N:\sigma$. The same holds when the first hypothesis is a multi-type judgment.
\end{lemma}

	\begin{proof}
	By induction on the first hypothesis.
	
	\begin{itemize}
	\item If the last rule is \drule{ax} and $M=x$, then $\ms m=[\gr{\sigma}{\1}]\+\ms m'$ for some $\ms m'$. By \cref{lem:derivations-additive-refinement}, write $\Delta=\Delta_0\+\Delta_1$ with $\derives\Delta_0\vdash N:[\gr{\sigma}{\1}]$. By \cref{lem:derivations-discreteness}, $\derives\Delta_0\vdash N:\sigma$. By \thmitemref{lem:weakening:weakening}, \ie weakening, we can conclude.
	
	\item If the last rule is \drule{ax} and $M=y\neq x$, replace $\Gamma \+ (y:\ms n), x:\ms m$ with $\Gamma \+ \Delta \+ (y:\ms n)$ in the context of the rule.
	
	\item If the last rule is \drule{\lto_i}, $\sigma = \ms n \lto \tau$, and $M=\lambda y.P$ (choosing $y\notin\set{x}\cup\fv(N)\cup\support(\Delta)$), apply induction to the premise, then apply the rule again:
		\[
			\begin{prooftree}
			\hypo{\Gamma,y:\ms n,x:\ms m\vdash P:\tau}
			\infer[dashed]1[\text{I.H.}]{\Gamma\+\Delta,y:\ms n\vdash P\subst N:\tau}
			\infer1[\lto_i]{\Gamma\+\Delta\vdash\lambda y.P\subst N:\ms n\lto\tau}.
			\end{prooftree}
		\]
		
	\item If the last rule is \drule{\lto_e} and $M=PQ$, its premises have contexts $\Gamma_P,x:\ms m_P$ and $\Gamma_Q,x:\ms m_Q$, where $\Gamma=\Gamma_P\+\Gamma_Q$ and $\ms m=\ms m_P\+\ms m_Q$. By \cref{lem:derivations-additive-refinement}, write $\Delta=\Delta_P\+\Delta_Q$ with $\derives\Delta_P\vdash N:\ms m_P$ and $\derives\Delta_Q\vdash N:\ms m_Q$. Apply induction and form
		\[
			\begin{prooftree}
			\hypo{\Gamma_P\+\Delta_P\vdash P\subst N:\ms n\lto\sigma}
			\hypo{\Gamma_Q\+\Delta_Q\vdash Q\subst N:\ms n}
			\infer2[\lto_e]{\Gamma\+\Delta\vdash(PQ)\subst N:\sigma}.
			\end{prooftree}
		\]
		
	\item If the last rule is
	\[
		\begin{prooftree}
		\hypo{[\gr{(\Gamma_k,x:\ms m_k\vdash M:\tau_k)}{a_k}]_{k\in K}}
		\infer1[!]{\Gamma,x:\ms m\vdash M:\ms n},
		\end{prooftree}
	\]
	where $\Gamma=\SUM_{k\in K}a_k\*\Gamma_k$, $\ms m=\SUM_{k\in K}a_k\*\ms m_k$, and $\ms n=[\gr{\tau_k}{a_k}]_{k\in K}$, then by \cref{lem:derivations-mixed-refinement}, there are $a_{kl}$ and $\Delta_{kl}$ such that
	\[
		\forall k\in K,\ a_k=\SUM_{l\in L_k}a_{kl},\qquad
		\Delta=\SUM_{k,l}a_{kl}\*\Delta_{kl},\qquad
		\forall k\in K,l\in L_k,\ \derives\Delta_{kl}\vdash N:\ms m_k.
	\]
	Apply induction to each premise $\derives\Gamma_k,x:\ms m_k\vdash M:\tau_k$, using $\derives\Delta_{kl}\vdash N:\ms m_k$ as the second hypothesis, and form the rule \drule{!} again:
	\[
		\begin{prooftree}
		\hypo{[\gr{(\Gamma_k\+\Delta_{kl}\vdash M\subst N:\tau_k)}{a_{kl}}]_{k\in K,l\in L_k}}
		\infer1[!]{\Gamma\+\Delta\vdash M\subst N:\ms n},
		\end{prooftree}
	\]
	since $\SUM_{k,l}a_{kl}\*(\Gamma_k\+\Delta_{kl})=\Gamma\+\Delta$ and $[\gr{\tau_k}{a_{kl}}]_{k,l}=[\gr{\tau_k}{a_k}]_k$.
	\qedhere
	\end{itemize}
	\end{proof}

\begin{lemma}[Backward Substitution]
	\label{lem:backward-substitution}
	Let $x\notin\fv(N)\cup\support(\Gamma)$. If $\derives\Gamma\vdash M\subst N:\sigma$, then there are $\Gamma_M$, $\Gamma_N$ and $\ms m$ such that
	\[
		\Gamma=\Gamma_M\+\Gamma_N,\qquad
		\derives\Gamma_M,x:\ms m\vdash M:\sigma,\qquad
		\derives\Gamma_N\vdash N:\ms m,
	\]
	with $\Gamma_M(x)=\Gamma_N(x)=[]$. The same holds for multi-type judgments.
\end{lemma}

	\begin{proof}
	By induction on the input derivation and case distinction on $M$.
	
	\begin{itemize}
	\item If $M=x$ and the conclusion has an ordinary type $\sigma$, take $\ms m=[\gr{\sigma}{\1}]$, $\Gamma_M=\emptyset$, $\Gamma_N=\Gamma$, and form
		\[
			\begin{prooftree}[center=false]
			\infer0[ax]{x:[\gr{\sigma}{\1}]\vdash x:\sigma}
			\end{prooftree}
			\qquad
			\begin{prooftree}[center=false]
			\hypo{\Gamma\vdash N:\sigma}
			\infer1[!]{\Gamma\vdash N:[\gr{\sigma}{\1}]}.
			\end{prooftree}
		\]
		
	\item If $M=y\neq x$ and the last rule is \drule{ax}, take $\Gamma_M=\Gamma$, $\Gamma_N=\emptyset$, $\ms m=[]$, and use the nullary \drule{!} rule to type $N$.
	
	\item If $M=\lambda y.P$ (choosing $y \notin \set x \cup \fv(N)$), the last rule is \drule{\lto_i}, and $\sigma = \ms n \lto \tau$, apply induction to the premise $\derives \Gamma, y:\ms n \vdash P \subst N : \tau$ and obtain:
	\[	\Gamma = \Gamma_P \+ \Gamma_N \quad 
		\ms n = \ms n_P \+ \ms n_N \quad
		\derives \Gamma_P, y:\ms n_P, x:\ms m \vdash P:\tau \quad
		\derives \Gamma_N, y:\ms n_N \vdash N:\ms m. \]
	Apply respectively \thmitemref{lem:weakening:weakening} and \thmitemref{lem:weakening:remove-useless-variables} to the derivations for $P$ and $N$, obtaining $\derives \Gamma_P, y:\ms n, x:\ms m \vdash P:\tau$ and $\derives \Gamma_N \vdash N:\ms m$. Conclude by forming the rule \drule{\lto_i} again.
	
	\item If $M=PQ$ and the last rule is
	\[	\begin{prooftree}
		\hypo{\Gamma_P \vdash P\subst N:\ms n \lto \sigma}
		\hypo{\Gamma_Q \vdash Q\subst N:\ms n}
		\infer2[\lto_e]{\Gamma \vdash (PQ)\subst N:\sigma,}
		\end{prooftree} \]
	first observe that $(\Gamma_P \+ \Gamma_Q)(x) = \Gamma(x) = []$, so either by positivity (if $\Semiring$ has it) or by \thmitemref{lem:weakening:remove-useless-variables} we can assume that $\Gamma_P(x) = \Gamma_Q(x) = []$. Induction gives
	\[
		\begin{prooftree}
		\hypo{\Gamma_P,x:\ms m_P\vdash P:\ms n\lto\sigma}
		\hypo{\Gamma_Q,x:\ms m_Q\vdash Q:\ms n}
		\infer2[\lto_e]{\Gamma_P\+\Gamma_Q,x:\ms m_P\+\ms m_Q\vdash PQ:\sigma},
		\end{prooftree}
	\]
	and $\derives\Gamma_{N,R}\vdash N:\ms m_R$ for $R\in\set{P,Q}$, with $\Gamma=\Gamma_P\+\Gamma_Q\+\Gamma_{N,P}\+\Gamma_{N,Q}$. The converse implication of \cref{lem:derivations-additive-refinement} gives $\derives\Gamma_{N,P}\+\Gamma_{N,Q}\vdash N:\ms m_P\+\ms m_Q$.

	\item Otherwise, if the last rule is
	\[	\begin{prooftree}
		\hypo{[\gr{(\Gamma_i \vdash M \subst N:\tau_i)}{c_i}]_{i\in I}}
		\infer1[!]{\Gamma\vdash M\subst N:\ms n},
		\end{prooftree} \]
	where $\Gamma=\SUM_{i\in I}c_i\*\Gamma_i$ and $\ms n=[\gr{\tau_i}{c_i}]_{i\in I}$,
	then first \thmitemref{lem:weakening:remove-useless-variables} in each hypothesis as in the previous case. Then by induction, for all $i\in I$, 
	$\Gamma_i=\Gamma_{M,i}\+\Gamma_{N,i}$ with $\derives\Gamma_{M,i},x:\ms m_i\vdash M:\tau_i$, and $\derives\Gamma_{N,i}\vdash N:\ms m_i$. Put
	\[
		\Gamma_M\eqdef\SUM_{i\in I}c_i\*\Gamma_{M,i},\qquad
		\Gamma_N\eqdef\SUM_{i\in I}c_i\*\Gamma_{N,i},\qquad
		\ms m\eqdef\SUM_{i\in I}c_i\*\ms m_i.
	\]
	The first desired derivation is form using the rule \drule{!} again:
	\[
		\begin{prooftree}
		\hypo{[\gr{(\Gamma_{M,i},x:\ms m_i\vdash M:\tau_i)}{c_i}]_{i\in I}}
		\infer1[!]{\Gamma_M,x:\ms m\vdash M:[\gr{\tau_i}{c_i}]_{i\in I}}.
		\end{prooftree}
	\]
	The second one can be obtained using the converse implication of \cref{lem:derivations-mixed-refinement}, giving $\derives\Gamma_N\vdash N:\ms m$. Distributivity gives $\Gamma_M\+\Gamma_N=\SUM_{i\in I}c_i\*\Gamma_i=\Gamma$.
	\qedhere
	\end{itemize}
	\end{proof}

\begin{theorem}[Subject Reduction and Subject Expansion]
	\label{the:srse-finite-multiset}
	For all weak resource semiring~$\Semiring$, $\itypes$ enjoys subject reduction and subject expansion, \ie, for all $M\bred N$, $\Gamma$ and $\sigma$,
	\[
		\derives\Gamma\vdash M:\sigma
		\quad\text{\ifandonlyif}\quad
		\derives\Gamma\vdash N:\sigma.
	\]
	The same holds for multi-type judgments.
\end{theorem}

	\begin{proof}
	By induction on $M \bred N$. In the base case, \ie $(\lambda x.P)Q\bred P\subst Q$, apply \cref{lem:forward-substitution,lem:backward-substitution}. The induction cases are straightforward.
	\end{proof}

In particular $\itypes[\Bool]$, $\itypes[\Nat]$ and $\itypes[\Tropical]$ enjoy subject reduction and expansion.

\subsection{Extension to the Relevant System $\ritypes$}
\label{sec:sr-se:relevant}

Weak discreteness does not suffice for relevant subject reduction. For example, using the semiring $\Bool$, put $\tau_0=[]\lto\alpha$, $\tau_1=[\gr\alpha\1]\lto\alpha$, $\Delta_0=y:[\gr{\tau_0}\1]$ and $\Delta_1=y:[\gr{\tau_1}\1],z:[\gr\alpha\1]$. We have $\derives\Delta_i\vdash yz:\alpha$ for $i=0,1$, hence
\[
	\begin{prooftree}
	\hypo{\emptyset\vdash\lambda x.x:[\gr\alpha\1]\lto\alpha}
	\hypo{\Delta_0\vdash yz:\alpha}
	\hypo{\Delta_1\vdash yz:\alpha}
	\infer2[!]{\Delta_0\+\Delta_1\vdash yz:[\gr\alpha\1]}
	\infer2[\lto_e]{\Delta_0\+\Delta_1\vdash(\lambda x.x)(yz):\alpha}.
	\end{prooftree}
\]
But $\Delta_0\+\Delta_1\vdash yz:\alpha$ is not derivable: the single occurrence of the head variable uses only one of $\tau_0,\tau_1$, and the argument $z$ cannot consume the other entry at $y$. To remove this obstruction, we need to consider a \emph{strong} resource semiring.

\begin{theorem}
	\label{the:srse-relevant}
	For all strong resource semiring $\Semiring$, the relevant system $\ritypes$ enjoys subject reduction and subject expansion.
\end{theorem}

	\begin{proof}
	The proof is as for \cref{the:srse-finite-multiset}, except for the following changes.
	
	In the proof of forward substitution (\cref{lem:forward-substitution}):
	\begin{itemize}
	\item For the case $M = x$, use the \drule{rax} rule and the strongly discrete version of \cref{lem:derivations-discreteness}.
	\item For the case $M = y \neq x$, the rule \drule{rax} implies that $\ms m = []$, and the 0-ary instance of \cref{lem:derivations-additive-refinement} gives $\Delta = \emptyset$, so that we can just keep the initial derivation untouched.
	\end{itemize}
	
	In the proof of backward substitution (\cref{lem:backward-substitution}):
	\begin{itemize}
	\item The two variable cases use the rule \drule{rax}.
	\item In the abstraction case, instead of applying \thmitemref{lem:weakening:weakening} and \thmitemref{lem:weakening:remove-useless-variables} to manipulate the derivations, use \thmitemref{lem:weakening:relevance} to show that no manipulation is needed.
	\qedhere
	\end{itemize}
	\end{proof}

In particular $\ritypes[\Nat]$ enjoys subject reduction and expansion.

\section{An Extension to Infinitary Types}
\label{sec:lf}

\subsection{Locally Finite Sums and Multisets}

\begin{definition}[Topological Semiring]
	\label{def:topological-semiring}
	A \emph{topological semiring} is a semiring $\Semiring$ equipped with a Hausdorff topology making both addition and multiplication continuous.
\end{definition}

We give $\Nat$ and $\Bool$ the discrete topology. On $\Tropical$ we consider
the topology induced by the ultrametric
\[	\dist[\Tropical](a,b) \eqdef \begin{cases}
	0 & \text{if $a=b$,} \\ 
	2^{-\min(a,b)} & \text{otherwise.}
	\end{cases} \]

We use the topology on $\Semiring$ to accommodate infinite sums in a controlled way, namely by imposing a condition of \emph{local finiteness}.

\begin{definition}[Locally Finite Families and Sums]
\label{def:lf-families-sums}
	A family $(a_i)_{i\in I}$ in a topological semiring $\Semiring$ is \emph{locally finite} if for all neighbourhood $U$ of $\0$, the set $\set{i\in I}[a_i\notin U]$ is finite.
	Its sum, when it exists, is
	\[
		\SUM_{i\in I}a_i\eqdef
		\lim_{F\subsetf I}\SUM_{i\in F}a_i,
	\]
	where $F\subsetf I$ means that $F$ is a finite subset of $I$, hence the limit ranges over finite subsets directed by inclusion. We say that $\Semiring$ \emph{admits locally finite sums} if:
	\begin{enumerate}
		\item every locally finite family has a sum;
		\item for all locally finite family $(a_i)_{i\in I}$ and partition $I=\bigsqcup_{j\in J}I_j$,
		the family $\left(\SUM_{i\in I_j}a_i\right)_{j \in J}$ is locally finite
		and $\SUM_{i\in I}a_i=\SUM_{j\in J}\SUM_{i\in I_j}a_i$;
		\thmitemlabel{def:lf-families-sums:partition}
		\item if $(a_i)_{i\in I}$ and each $(b_{ij})_{j\in J}$ are locally finite, then $(a_i\*b_{ij})_{i\in I,j\in J}$ is locally finite.
		\thmitemlabel{def:lf-families-sums:product}
	\end{enumerate}
\end{definition}

Hausdorffness gives uniqueness of the sum of a family: two distinct limits would have disjoint neighbourhoods, whereas convergence would force every sufficiently late finite partial sum to belong to both.

A finitely supported family is locally finite; the converse holds for the discrete topology. In particular, this construction is of no consequence for $\Nat$ and $\Bool$.

Continuity of the semiring operations and Item~3 in the above Definition give distributivity over locally finite sums:
$a\*\SUM_i b_i=\SUM_i a\*b_i$ (and similarly on the right side).

\begin{definition}[Locally Finite Resource Semiring]
\label{def:lf-resource-semiring}
	Given a semiring $\Semiring$ admitting locally finite sums,
	consider the following conditions:
	\begin{itemize}
	\item \emph{Locally finite discreteness}:
		For all locally finite family $(a_i)_{i \in I}$,
		if $\SUM_{i\in I} a_i=\1$ then exactly one $a_i$ is nonzero (and is equal to~$\1$).
	\item \emph{Locally finite weak discreteness}:
		For all locally finite family $(a_i)_{i \in I}$,
		if $\SUM_{i\in I} a_i=\1$ then $\exists i\in I,\ a_i \>= \1$.
	\item \emph{Locally finite additive refinement}:
		For all locally finite families $(a_i)_{i\in I}$, $(b_j)_{j\in J}$,
		and labellings $\lambda : I \to X$, $\mu : J \to X$,
		\begin{multline*}
		\left( \forall x\in X,\ 
		\SUM_{\substack{i\in I \\ \lambda(i)=x}} a_i
			= \SUM_{\substack{j\in J \\ \mu(j)=x}} b_j \right)
		\implies \exists (c_{ij})_{i\in I,j\in J} \text{ locally finite},\\
		\forall i\in I,\ a_i=\SUM_{j\in J} c_{ij},\quad
		\forall j\in J,\ b_j=\SUM_{i\in I} c_{ij},\quad
		\forall i\in I, j\in J,\ \lambda(i) \neq \mu(j) \implies c_{ij}=\0.
		\end{multline*}
	\item \emph{Simultaneous locally finite mixed refinement}:
		For all locally finite families $(a_k)_{k\in K}$, $(b_{xk})_{x\in X}$ for every $k \in K$, and $(c_i)_{i\in I}$,
		and for all labelling $\lambda:I\to X$,
		\begin{multline*}
			\begin{aligned}
			& \left( \forall x\in X,\
				\SUM_{k\in K} a_k\*b_{xk}
					= \SUM_{\substack{i\in I \\ \lambda(i)=x}} c_i
			\right) \implies \\
			&\begin{array}[t]{l}
				\exists (L_k)_{k\in K},\\
				\exists (a_{kl})_{k\in K, l\in L_k} \text{ locally finite},\\ 
				\exists (b_{ikl})_{i\in I, k\in K, l\in L_k} \text{ such that $(b_{ikl})_{i\in I}$ is locally finite for all $k \in K$ and $l \in L_k$},
			\end{array}
			\end{aligned} \\
			\begin{gathered}
			\forall k\in K,\ a_k=\SUM_{l\in L_k} a_{kl},\qquad
			\forall x\in X,k\in K,l\in L_k,\
				b_{xk}=\SUM_{\substack{i\in I \\ \lambda(i)=x}} b_{ikl}, \\
			\forall i\in I,\
				c_i=\SUM_{\substack{k\in K \\ l\in L_k}} a_{kl}\*b_{ikl}.
			\end{gathered}
		\end{multline*}
	\end{itemize}
	Then $\Semiring$ is said to be a \emph{(weak) locally finite resource semiring} whenever it enjoys locally finite (weak) discreteness, locally finite additive refinement, and simultaneous locally finite mixed refinement.
\end{definition}

In general, these locally finite axioms \emph{do not imply the finite ones} from \cref{def:resource-semiring} and \cref{lem:nary}: under the hypotheses of finite mixed refinement, locally finite mixed refinement does not always produce a finite refining family. We will build an example in \cref{rem:finite-and-locally-finite-refinement}.
Nonetheless, when $\Semiring$ is equipped with the discrete topology, the finite and locally finite definitions coincide again.
In particular, every (weak) resource semiring equipped with the discrete topology is a (weak) locally finite resource semiring. This is the topology we use for $\Nat$ and $\Bool$.

Observe that in a weak locally finite resource semiring, positivity and absence of zero divisors are consequences of additive and mixed refinement, respectively. One can actually get a locally finite versions of the former, namely $\SUM_i a_i=\0\implies\forall i,\ a_i=\0$, by applying additive refinement to the empty family.

\begin{lemma}
	\label{lem:tropical-local}
	$\Tropical$ is a weak locally finite resource semiring.
\end{lemma}
\begin{proof}
	A neighbourhood basis of $\infty$ is $U_d\eqdef\set{\infty}\cup\set{n\in\natset}[n>d]$. Thus $(a_i)_{i\in I}$ is locally finite \ifandonlyif $\set{i\in I}[a_i \leq d]$ is finite for every $d$. Its sum is $\min_{i \in I} a_i$. Using these characterisations, checking that the requirements of \cref{def:lf-families-sums,def:lf-resource-semiring} hold are straightforward algebraic manipulations.
\end{proof}

\begin{definition}[Locally Finite $\Semiring$-Multisets]
	\label{def:lf-msets}
	An $\Semiring$-multiset $\ms m:X\to\Semiring$ is \emph{locally finite} if, for all neighbourhood $U$ of $\0$, the set $\set{x\in X}[\ms m(x)\notin U]$ is finite. We denote the set of locally finite $\Semiring$-multisets by $\lfmsets[\Semiring]{X}$.
	Given a family $(x_i)_{i\in I}$ of elements of $X$ and a locally finite family $(a_i)_{i\in I}$ of coefficients, we write $[\gr{x_i}{a_i}]_{i\in I}$ for the $\Semiring$-multiset $\ms m:x\mapsto\SUM_{i\in I,x_i=x}a_i$.
\end{definition}

Only locally finite multisets occur in the following: we do not consider unrestricted maps $X\to\Semiring$. For $\Nat$ and $\Bool$, local finiteness and finiteness coincide, so that $\fmsets[\Nat]{X}$ and $\lfmsets[\Nat]{X}$ (\resp $\fmsets[\Bool]{X}$ and $\lfmsets[\Bool]{X}$) are both equal to the set of finite multisets (\resp finite sets) of elements of $X$. For $\Tropical$, a map $\ms m:X\to\Tropical$ is locally finite whenever
\[
	\forall d\in\natset,\qquad
	\set{x\in X}[\ms m(x)\leq d]\text{ is finite}.
\]

If $\Semiring$ admits locally finite sums, the pointwise sum and external product from \cref{def:msets} preserve local finiteness; so do weighted sums $\SUM_{i\in I}a_i\*\ms m_i$ with $(a_i)_{i\in I}$ locally finite.

\subsection{Locally Finite Intersection Types}

We now allow types to be infinite in two different directions:
\begin{itemize}
\item the $\Semiring$-multisets involved may be infinite: this is controlled by local finiteness, as introduced in the previous Subsection;
\item the syntax tree of the types may be infinitely deep, \ie coinductive (which is not the case for standard intersection types, but has already been considered in the literature \autocite{Vial.17,Vial.21,Cerda.Dal.26}): this kind of infinity is controlled by a mechanism inspired by guarded recursion, namely, coinductive recursion is allowed in an element $\sigma$ of an $\Semiring$-multiset of types $\ms m$ whenever its grade $\ms m(\sigma)$ is \enquote{guarding}, which motivates the following Definition.
\end{itemize}

\begin{definition}[Guarding Set]
\label{def:guarding-set}
	A \emph{guarding set} of $\Semiring$ is a subset $G\subseteq\Semiring$ such that, for all $a,b\in\Semiring$,
	\[
		\0\in G,\quad \1\notin G,\qquad
		a\+b\in G\iff a,b\in G,\qquad
		a\*b\in G\iff a\in G\text{ or }b\in G.
	\]
	Equivalently, $G=\gamma^{-1}(\set{0})$ for a unital semiring morphism $\gamma:\Semiring\to\Bool$; this morphism is uniquely determined by $G$.
	We additionally assume that $\gamma$ is continuous, or equivalently that $G$ is clopen.
\end{definition}

We take $G=\set{0}$ for $\Nat$ and $\Bool$; more generally we can do so for any resource semiring equipped with the discrete topology. For $\Tropical$ we take $G=\set{1,2,\dots,\infty}$.

\begin{definition}[Locally Finite $\Semiring$-Intersection Types]
	\label{def:lfitypes}
	Let $\Semiring$ be a semiring admitting locally finite sums,
	equipped with a guarding set $G$,
	and let $\Atoms$ be a countable set of atoms.
	The sets $\lfitypes[\Semiring]$ of \emph{locally finite $\Semiring$-intersection types} and $\lfitypes![\Semiring] \subset \lfmsets[\Semiring]{\lfitypes[\Semiring]}$ of \emph{locally finite $\Semiring$-intersection multi-types} are defined by mutual induction and coinduction by the following system of formation rules:
	\begin{gather*}
	\begin{prooftree}
		\hypo{ α \in \Atoms }
		\infer1[\Atoms]{ α \in \lfitypes[\Semiring] }
	\end{prooftree}
	\qquad\qquad
	\begin{prooftree}
		\hypo{ \ms{m} \in \lfitypes![\Semiring] }
		\hypo{ \tau \in \lfitypes[\Semiring] }
		\infer2[\lto]{ \ms{m} \lto \tau \in \lfitypes[\Semiring] }
	\end{prooftree}
	\\[\topsep]
	\begin{prooftree}[center]
		\hypo{ \left[ \gr {(\sigma_i \inlater{a_i} \lfitypes[\Semiring])}
			{a_i} \right]_{i \in I} }
		\infer1[!]{ [\gr{\sigma_i}{a_i}]_{i \in I} \in \lfitypes![\Semiring] }
	\end{prooftree}
	\qquad
	\begin{prooftree}[center]
		\hypo{ \sigma \in \lfitypes[\Semiring] }
		\hypo{ a \in G }
		\infer[double]2[coind]
		{ \sigma \inlater{a} \lfitypes[\Semiring] }
	\end{prooftree}
	\qquad
	\begin{prooftree}[center]
		\hypo{ \sigma \in \lfitypes[\Semiring] }
		\infer1[ind]{ \sigma \inlater{a} \lfitypes[\Semiring] }
	\end{prooftree}
	\end{gather*}
	where single-bar rules are treated inductively whereas double-bar rules are treated coinductively: concretely, every infinite branch must traverse infinitely many instances of \drule{coind}. Notice again that the formation rule \drule{!} has a (locally finite) $\Semiring$-multiset of hypotheses.
\end{definition}

Typing judgements keep the same shape as in the finite system, with contexts being finitely supported maps $\Var \to \lfitypes![\Semiring]$.

\begin{definition}[Typing] \label{def:lfitypes-typing}
	A λ-term $M$ is \emph{typable in $\lfitypes[\Semiring]$} whenever there is a derivation $\derivD$ with conclusion $\Gamma \vdash M : \sigma$ in the following (inductive) system of typing rules:
	\begin{gather*}
		\begin{prooftree}
		\infer0[ax]{ \Gamma \+ x:[\gr{\sigma}{\1}] \vdash x:\sigma }
		\end{prooftree}
		\qquad\qquad
		\begin{prooftree}
		\hypo{ \Gamma, x:\ms{m} \vdash M:\tau }
		\infer1[\lto_i]{ \Gamma \vdash λx.M : \ms{m} \lto \tau }
		\end{prooftree}
		\\[\topsep]
		\begin{prooftree}
		\hypo{ \Gamma \vdash M : \ms{m} \lto \tau }
		\hypo{ \Gamma' \vdash N : \ms{m} }
		\infer2[\lto_e]{ \Gamma \+ \Gamma' \vdash MN : \tau }
		\end{prooftree}
		\qquad
		\begin{prooftree}
		\hypo{ \left[ \gr {(\Gamma_i \vdash N:\sigma_i)} {a_i} \right]
			_{i \in I} }
		\infer1[!]{ \SUM_{i \in I} a_i \* \Gamma_i
			\vdash N :
			[\gr{\sigma_i}{a_i}]_{i \in I}
		}
		\end{prooftree}
	\end{gather*}
	for some context $\Gamma$ and $\Semiring$-intersection type $\sigma$.
	We then write $\derivD \derives \Gamma \vdash M : \sigma$, and simply $\derives \Gamma \vdash M : \sigma$ to say that such a $\derivD$ exists.
\end{definition}

A look at this typing system motivates our restriction to guarded corecursive types: removing the guard makes the typing unsound in general. For instance in $\lfitypes[\Bool]$, the unguarded definition $\tau \eqdef [\gr\tau1]\lto\alpha$ makes $\lambda x.xx$ typable at $\tau$ and hence makes $\Omega \eqdef (\lambda x.xx)(\lambda x.xx)$ typable at $\alpha$.

The relevant system $\lfritypes$ is obtained by replacing \drule{ax} with \drule{rax}, as in the finite system. The weakening and relevance properties of \cref{lem:weakening} extend to these systems, by the same proofs.

The full system $\lfitypes[\Tropical]$ is the tropical intersection type system that we previously studied \cite{Cerda.Dal.26}.

\begin{theorem}
	\label{the:srse-locally-finite}
	For all weak locally finite resource semiring $\Semiring$, the full system $\lfitypes$ enjoys subject reduction and subject expansion.
\end{theorem}

	\begin{proof}
	We extend the three lemmas of \cref{sec:sr-se:algebraic} as follows.
	
	\begin{itemize}
	\item In \cref{lem:derivations-discreteness}, locally finite weak discreteness plays the role of $n$-ary weak discreteness.
	
	\item In \cref{lem:derivations-additive-refinement}, keep $J$ finite. In the decomposition of the last rule \drule{!}, the index set $I$ becomes infinite, with a locally finite family $(a_i)_{i\in I}$. The family $(\ms m_j(\sigma))_{(j,\sigma)\in J\times\lfitypes}$ is locally finite by \cref{def:lf-msets}. Apply locally finite additive refinement to these two families, with labellings
		\[
			\lambda:I\to\lfitypes,\ i\mapsto\sigma_i,\qquad
			\mu:J\times\lfitypes\to\lfitypes,\ (j,\sigma)\mapsto\sigma.
		\]
	The resulting $c_{i,(j,\sigma)}$ are equal to~$\0$ unless $\sigma=\sigma_i$; put $c_{ij}\eqdef c_{i,(j,\sigma_i)}$, which defines a locally finite family.
		
	\item In \cref{lem:derivations-mixed-refinement}, take $(a_k)_{k\in K}$ locally finite and each $\ms m_k$ locally finite. Write the given rule~\drule{!} with locally finite coefficients $(c_i)_{i\in I}$ and premises $\Delta_i\vdash N:\sigma_i$. Put $X\eqdef\lfitypes$ and $\lambda:I\to X,\ i\mapsto\sigma_i$;
	observe that $(a_k)_{k\in K}$, each $(\ms m_k(\sigma))_{\sigma\in X}$, and $(c_i)_{i\in I}$ are locally finite families, and that
	\[
		\forall\sigma\in X,\ \SUM_{k\in K}a_k\*\ms m_k(\sigma)
		=\SUM_{\substack{i\in I\\\lambda(i)=\sigma}}c_i.
	\]
	Apply simultaneous locally finite mixed refinement. It gives $(L_k)_{k\in K}$, a locally finite family $(a_{kl})_{k\in K,l\in L_k}$, and families $(b_{ikl})_{i\in I}$ locally finite for every $k,l$, with
	\begin{gather*}
		\forall k,\ a_k=\SUM_{l\in L_k}a_{kl},\qquad
		\forall\sigma,k,\ \forall l\in L_k,\ \ms m_k(\sigma)=\SUM_{\substack{i\in I\\\sigma_i=\sigma}}b_{ikl},\qquad
		\forall i,\ c_i=\SUM_{k,l}a_{kl}\*b_{ikl}.
	\end{gather*}
	The conclusion of the proof is as in the finite case.
	\end{itemize}
	
	Once these adaptions are performed, the proof of forward and backward substitution are exactly the same; all flattenings and regroupings in these proofs are justified by \cref{def:lf-families-sums}; positivity and absence of zero divisors in $\Semiring$ are needed to justify that forward and backward substitution preserve finitely supported typing contexts.
	\end{proof}

In particular $\lfitypes[\Tropical]$ enjoys subject reduction and expansion.

\begin{theorem}
	\label{the:lf-srse-relevant}
	For all (strong) locally finite resource semiring $\Semiring$, the relevant system $\lfritypes$ enjoys subject reduction and subject expansion.
\end{theorem}

	\begin{proof}
	Apply the proof of \cref{the:srse-locally-finite} with the relevant axiom. The changes in the substitution lemmas are those of \cref{the:srse-relevant}; locally finite discreteness replaces $n$-ary discreteness in the proof of \cref{lem:derivations-discreteness}. The converse constructions still require only the semiring laws and the properties of locally finite sums.
	\end{proof}

\begin{remark}[Intermediary Fragments]
	\label{rem:intermediate-fragments}
	One can also consider some fragments lying between the finite system $\itypes$ and the locally finite system $\lfitypes$.
	
	\begin{itemize}
	\item The \emph{finite-multiset fragment} requires finite support for every multiset, in annotations and in the hypotheses of the formation and typing rules \drule{!}.
	\item The \emph{inductive fragment} of $\lfitypes$ uses types and multi-types formed without \drule{coind}, \ie, with well-founded formation trees.
	\end{itemize}
	
	Subject reduction and expansion hold in the finite-multiset (\resp inductive) fragment under the hypotheses of \cref{the:srse-finite-multiset} (\resp \cref{the:srse-locally-finite}), \ie if $\Semiring$ is a weak resource semiring (\resp weak locally finite resource semiring).
	
	The intersection of the finite-multiset and inductive fragments is the finite system $\itypes$ from \cref{sec:finite:typing}.
	The same restrictions define fragments of $\lfritypes$, whose intersection is $\ritypes$.
\end{remark}

\section{The Multiset Lifting}
\label{sec:multiset-constr}

In this section we describe the following construction: instead of taking coefficients in $\Semiring$ as grades (as in $\itypes$ and $\lfitypes$), we take \emph{multisets of (non-zero) coefficients}: whenever a same abstracted variable $x$ (in some term $λx.M$) has occurrences typed by $\sigma$ with grades, say, $a$ and $b$, the type of $λx.M$ is not any more $[\gr{\sigma}{a\+b}] \lto \tau$ but $[\gr{\sigma}{[a,b]}] \lto \tau$. This is reminiscent of the shift from idempotent to non-idempotent intersection types; actually it is an instance of the construction (for $\Semiring = \Bool$, see \cref{exa:multiset-constr-boolean} below).

\subsection{The Finite Lifting of the Finite System}
\label{sec:multiset-constr:finite}

In the finite case, the construction relies on the following observation, already made in \autocite[Proposition~3]{Breuvart.Pag.15}.

\begin{lemma}
	\label{def:semiring-of-msets}
	\label{lem:monoid}
	For every monoid $(\Monoid,\times,\1)$, equipping $\fmsets{\Monoid}$ with the following operations:
	\[
		(\ms a\msplus\ms b)(c)\eqdef\ms a(c)+\ms b(c),\qquad
		(\ms a\mstimes\ms b)(c)\eqdef\SUM_{\substack{a,b\in\Monoid\\a\* b=c}}\ms a(a)\ms b(b),
	\]
	with respective neutral elements $[]$ and $[\1]$, defines a resource semiring.
\end{lemma}

	\begin{proof}
	The support of a sum is included in the union of the supports, and the support of a product is included in their pointwise product. Thus both operations preserve finite support. The semiring laws follow from the monoid laws and the semiring laws on $\Nat$.
	
	Positivity, absence of zero divisors, discreteness and additive refinement follow from the pointwise application of the corresponding properties in $\Nat$. 
	For mixed refinement, suppose that $\ms a\mstimes\ms b=\ms c_1\msplus\ms c_2$. Write $\ms a=[a_l]_{l\in L}$ with $L$ finite. For each $c\in\Monoid$, put $J_c\eqdef\set{(l,b)}[l\in L,\ b\in\support(\ms b),\ a_l\* b=c]$. The hypothesis gives
	\[
		\SUM_{(l,b)\in J_c}\ms b(b)=\ms c_1(c)+\ms c_2(c).
	\]
	Apply $n$-ary additive refinement in $\Nat$. For each $(l,b)\in J_c$ it gives coefficients $b_{1l}(b)$, $b_{2l}(b)$ such that
	\[
		\ms b(b)=b_{1l}(b)+b_{2l}(b),\qquad
		\forall i\in\set{1,2},\ \ms c_i(c)=\SUM_{(l,b)\in J_c}b_{il}(b).
	\]
	Every pair $(l,b)\in L\times\support(\ms b)$ belongs to exactly one $J_c$, so these choices define $\ms b_{il}(b)\eqdef b_{il}(b)$ unambiguously; put $\ms b_{il}(b)\eqdef0$ for $b\notin\support(\ms b)$. Each $\ms b_{il}$ has finite support, and
	\[
		\ms a=\mssum_{l\in L}[a_l],\qquad
		\forall l\in L,\ \ms b=\ms b_{1l}\msplus\ms b_{2l},\qquad
		\forall i\in\set{1,2},\ \ms c_i=\mssum_{l\in L}[a_l]\mstimes\ms b_{il}.
		\qedhere
	\]
	\end{proof}

Observe that $\msplus$ is just multiset union, and the convolution product $\mstimes$ gives pointwise product of multiset elements: $[a_l]_l\mstimes[b_r]_r=[a_l\*b_r]_{l,r}$.

Let $\Semiring$ be a semiring with no zero divisors. Its nonzero elements form a multiplicative monoid $\Semiring*\eqdef(\Semiring\setminus\set{\0},\*,\1)$.

\begin{definition}[Finite Multiset Lifting]
	\label{def:finite-msgrades}
	We put $\fmsgrades\eqdef\fmsets{\Semiring*}$, with the sum and product of \cref{lem:monoid}. We denote multisets of grades by $\ms a$, $\ms b$, etc., whereas multisets of types (\eg,~in~$\itypes[\fmsgrades]$) are still denoted by $\ms m$,~$\ms n$,~etc.
\end{definition}

\begin{remark}
	\label{rem:finite-currying}
	For every set $X$, currying identifies $\fmsets[\fmsgrades]{X}$ with $\fmsets{X\times\Semiring*}$. Indeed, a finite union of finite supports is finite, and both projections of a finite set are finite. We denote the curried map by $\mscurr m$, with $\mscurr m(x,c)\eqdef\ms m(x)(c)$.
\end{remark}

\begin{corollary}[of \cref{lem:monoid}]
	\label{the:finite-msgrades-is-a-resource-semiring}
	$\fmsgrades$ is a resource semiring.
\end{corollary}

Consequently $\itypes[\fmsgrades]$ and $\ritypes[\fmsgrades]$ enjoy subject reduction and expansion, by \cref{the:srse-finite-multiset,the:srse-relevant}.

\begin{theorem}[Collapse]
	\label{the:finite-collapse}
	The \emph{collapse} map $\collapse:\fmsgrades\to\Semiring$ defined by
	\[
		\collapse:[a_l]_{l\in L}\longmapsto\SUM_{l\in L}a_l
	\]
	is a surjective semiring morphism. It extends to finite types and multi-types by
	\[
		\collapse(\alpha)\eqdef\alpha,\qquad
		\collapse(\ms m\lto\tau)\eqdef\collapse(\ms m)\lto\collapse(\tau),\qquad
		(\collapse(\ms m))(\tau)\eqdef\SUM_{\substack{\sigma\in\itypes[\fmsgrades]\\\collapse(\sigma)=\tau}}\collapse(\ms m(\sigma)),
	\]
	translating typings in $\itypes[\fmsgrades]$ into typings in $\itypes[\Semiring]$, and likewise for the relevant systems. If $\Semiring$ has positivity, this translation preserves positive typings.
\end{theorem}

	\begin{proof}
	Preservation of addition and multiplication are immediate. Surjectivity follows from $\collapse([c])=c$ for $c\neq\0[\Semiring]$, and $\collapse([])=\0[\Semiring]$.
	Induction on types and on typing derivations then proves that collapse translates valid typings into valid typings. If $\Semiring$ is positive, then $\collapse(\ms a)=\0[\Semiring]$ \ifandonlyif $\ms a=[]$. Collapse therefore preserves empty and nonempty multi-types, hence positive typings.
	\end{proof}

\begin{example}[Boolean Grades]
	\label{exa:multiset-constr-boolean}
	Since $\Bool*=\set{1}$, $\fmsgrades[\Bool]$ consists of finite multisets of $1$. The map $\ms a\mapsto\ms a(1)$ identifies it with $\Nat$. Consequently, $\itypes[\fmsgrades[\Bool]]$ is just the usual system $\itypes[\Nat]$, and likewise for the relevant systems. Collapse becomes the classical morphism turning non-idempotent into idempotent intersection typings.
\end{example}

\subsection{Lifting Locally Finite Types}

We now extend the construction to the locally finite setting from \cref{sec:lf}. Things get a bit more complicated: extending \cref{def:finite-msgrades} by taking grading multisets in $\lfmsets{\Semiring*}$ is not enough, since these are $\Nat$-multisets and $\Nat$ is equipped with the discrete topology, so that $\lfmsets{\Semiring*}=\fmsets{\Semiring*}$. We instead allow infinite multisets of grades whose family of elements is locally finite in $\Semiring$: outside every neighbourhood of $\0[\Semiring]$, only finitely many elements may occur.

Let $\Semiring$ admit locally finite sums and have no zero divisors. Its nonzero elements form a multiplicative monoid $\Semiring* \eqdef (\Semiring\setminus\set{\0}, \*, \1)$.

\begin{definition}
	\label{def:lfmsgrades}
	A multiset of grades $\ms a:\Semiring*\to\Nat$ has \emph{locally finite support towards $\0$} if, for all neighbourhood $U$ of $\0[\Semiring]$, the set $\support(\ms a)\setminus U$ is finite.
	We denote the set of these multisets by~$\lfmsgrades$.
	(Equivalently, $\lfmsgrades$ consists of multisets $[a_l]_{l\in L}$ whose family of elements $(a_l)_{l\in L}$ is locally finite in~$\Semiring$.)
	We equip $\lfmsgrades$ with:
	\begin{itemize}
	\item the topology whose neighbourhood basis at $\ms a$ consists of
	\[
		V_U(\ms a)\eqdef\set*{\ms b\in\lfmsgrades}[\forall c\in\Semiring*\setminus U,\ \ms b(c)=\ms a(c)],
	\]
	where $U$ ranges over the neighbourhoods of $\0[\Semiring]$;
	\item the sum and product defined as in \cref{def:finite-msgrades}; the sum is extended to any locally finite family $(\ms a_i)_{i\in I}$, by defining $\left(\mssum_{i\in I}\ms a_i\right)(c) \eqdef \SUM_{i\in I}\ms a_i(c)$.
	\end{itemize}
\end{definition}

To see that the product is well-defined, write $\ms a=[a_k]_{k\in K}$ and $\ms b=[b_l]_{l\in L}$. First, the families $(a_k)_k$ and $(b_l)_l$ are locally finite in $\Semiring$, so by \thmitemref{def:lf-families-sums:product} the family $(a_k\*b_l)_{k,l}$ is locally finite too. Since $\Semiring$ has no zero divisors, its elements belong to $\Semiring*$. Second, for each $c\in\Semiring*$, Hausdorffness gives a neighbourhood $U_c$ of $\0[\Semiring]$ not containing $c$. Hence $\set{k,l}[a_k \* b_l = c] \subseteq \set{k,l}[a_k \* b_l \notin U_c]$, which is finite by local finiteness of $(a_k\*b_l)_{k,l}$. As a consequence the sum defining $(\ms a\mstimes\ms b)(c)$ is a finite sum.

Local finiteness of $\ms a$ in this Definition concerns its family of elements; in particular it is \emph{not} the same thing as local finiteness of the multiset $\ms a$ as defined in \cref{def:lf-msets} (which just amounts to finite support, as we consider $\Nat$-multisets); neither does it require the coefficient family $(\ms a(c))_{c\in\Semiring*}$ to be locally finite in $\Nat$.

\begin{remark}
	\label{rem:currying}
	For every set $X$, currying identifies $\lfmsets[\lfmsgrades]{X}$ with the set of all maps $X\times\Semiring*\to\Nat$ such that, for all neighbourhood $U$ of $\0[\Semiring]$,
	\[
		\set*{(x,c)\in X\times(\Semiring*\setminus U)}[\ms m(x,c)\neq 0]\text{ is finite}.
	\]
	For $\ms m \in \lfmsets[\lfmsgrades]{X}$, we denote again the curried map by $\mscurr m$.
\end{remark}

\begin{theorem}
	\label{the:msgrades-is-a-resource-semiring}
	$\lfmsgrades$ is a locally finite resource semiring.
	In addition, if $G$ is a guarding set of $\Semiring$, then $\msgset G\eqdef\set{\ms a\in\lfmsgrades}[\support(\ms a)\subseteq G]$ is a guarding set of~$\lfmsgrades$.
\end{theorem}

	\begin{proof}
	We first check that \cref{def:lfmsgrades} defines a topological semiring (\cref{def:semiring,def:topological-semiring}).
	
	\begin{itemize}
	\item The semiring laws follow from the monoid laws on $\Semiring*$ and the semiring laws on $\Nat$.
	
	\item To see that the topology on $\lfmsgrades$ is Hausdorff, take $\ms a \neq \ms b$, then $\exists c\in\Semiring*,\ \ms a(c)\neq\ms b(c)$; any neighbourhood $U$ of $\0[\Semiring]$ not containing $c$ gives disjoint neighbourhoods $V_U(\ms a)$ and $V_U(\ms b)$.
	
	\item Continuity of addition is immediate.
	For continuity of multiplication at $(\ms a,\ms b)$, we first fix a neighbourhood $U$ of $\0[\Semiring]$ and we will find a neighbourhood $U'$ of $\0[\Semiring]$ such that
	\[
		V_{U'}(\ms a)\mstimes V_{U'}(\ms b)
		\subseteq V_U(\ms a\mstimes\ms b).
	\]
	Choose a neighbourhood $W$ of $\0[\Semiring]$ such that $W \*[\Semiring] W \subseteq U$. The sets $A\eqdef\support(\ms a)\setminus W$ and $B\eqdef\support(\ms b)\setminus W$ are finite. By continuity of multiplication in $\Semiring$, choose a neighbourhood $U'\subseteq W$ of $\0[\Semiring]$ such that
	\[
		\forall a\in A,\ a\*[\Semiring]U'\subseteq U,\qquad
		\forall b\in B,\ U'\*[\Semiring]b\subseteq U.
	\]
	If $a\in\support(\ms a)$, $b\in\support(\ms b)$ and $a\*[\Semiring]b\notin U$, then $a,b\notin U'$. Indeed, if $a\in U'$, then either $b\in W$, giving $a\*[\Semiring]b\in W\*[\Semiring]W\subseteq U$, or $b\in B$, giving $a\*[\Semiring]b\in U'\*[\Semiring]b\subseteq U$. The argument for $b\in U'$ is symmetric.
	
	Now take $\ms a'\in V_{U'}(\ms a)$ and $\ms b'\in V_{U'}(\ms b)$. For every $c\notin U$,
	\[
		(\ms a'\mstimes\ms b')(c)
		=
		\SUM_{\substack{
			a\in\support(\ms a)\setminus U'\\
			b\in\support(\ms b)\setminus U'\\
			a\*[\Semiring]b=c
		}}
		\ms a(a)\*[\Nat]\ms b(b)
		=
		(\ms a\mstimes\ms b)(c).
	\]
	This proves the required inclusion.
	\end{itemize}
	
	We now check that $\lfmsgrades$ admits locally finite sums (\cref{def:lf-families-sums}).
	
	\begin{enumerate}
	\item Let $(\ms a_i)_{i\in I}$ be locally finite, let us prove that it has a sum.
	By \cref{rem:currying}, for every neighbourhood $U$ of $\0[\Semiring]$, the set
	\[
		E_U\eqdef
		\set*{(i,c)\in I\times(\Semiring*\setminus U)}
		[\ms a_i(c) \neq 0]
	\]
	is finite. For fixed $c\in\Semiring*$ and any $U$ not containing $c$, we obtain that finitely many $\ms a_i(c)$ are non-zero. Thus
	$\ms a(c)\eqdef\SUM_{i\in I}\ms a_i(c)$
	is a finite sum in $\Nat$. In addition, for all neighbourhood $U$ of $\0[\Semiring]$, the set $\support(\ms a) \setminus U$ is included in the second projection of $E_U$, hence is finite, so that $\ms a \in \lfmsgrades$.
	To show that it is the sum of the family $(\ms a_i)_i$, observe that as soon as $F\subsetf I$ contains the first projection of $E_U$, then
	$\mssum_{i\in F}\ms a_i\in V_U(\ms a)$.
	
	\item Take a locally finite family $(\ms a_i)_{i \in I}$ and a partition $I=\bigsqcup_{j\in J}I_j$. For all $j \in J$ and $c \in \Semiring*$, if $\SUM_{i\in I_j}\ms a_i(c) \neq 0$, then $\ms a_i(c) \neq 0$ for some $i\in I_j$. As a consequence, for every neighbourhood $U$ of $\0[\Semiring]$,
	\[
		\set*{j\in J}[\mssum\nolimits_{i\in I_j}\ms a_i \notin V_U([])]
		\subseteq
		\set*{j\in J}[\exists(i,c)\in E_U,\ i\in I_j].
	\]
	The right-hand set is finite, since $E_U$ is finite and each $i$ belongs to exactly one $I_j$. Thus $(\mssum_{i\in I_j}\ms a_i)_{j\in J}$ is locally finite.
	Associativity and commutativity of the finite sums in $\Nat$, applied pointwise for every $c \in \Semiring*$, give $\mssum_{i\in I}\ms a_i = \mssum_{j\in J}\mssum_{i\in I_j}\ms a_i$.
	
	\item Let $(\ms a_i)_{i\in I}$ and each $(\ms b_{ij})_{j\in J}$ be locally finite. Write $\ms a_i=[a_{ik}]_{k\in K_i}$ and $\ms b_{ij}=[b_{ijl}]_{l\in L_{ij}}$. By \cref{rem:currying}, $(a_{ik})_{i,k}$ is locally finite in $\Semiring$, and so is $(b_{ijl})_{j,l}$ for every $i$. \thmitemref{def:lf-families-sums:product} gives local finiteness of
	$(a_{ik}\*[\Semiring]b_{ijl})_{i,j,k,l}$.
	Since $\ms a_i\mstimes\ms b_{ij} = [a_{ik}\*[\Semiring]b_{ijl}]_{k,l}$, \cref{rem:currying} gives local finiteness of $(\ms a_i\mstimes\ms b_{ij})_{i,j}$.
	\end{enumerate}
	
	Then we check algebraic axioms of a locally finite resource semiring (\cref{def:lf-resource-semiring}).
	
	\begin{itemize}
	\item For locally finite discreteness, suppose that $\mssum_i\ms a_i=[\1[\Semiring]]$. For all $c\neq\1[\Semiring]$, for all $i$, $\ms a_i(c)=0$ by positivity of $\Nat$; local finiteness of $(\ms a_i)_i$ and $n$-ary discreteness of $\Nat$ gives exactly one $\ms a_i(\1[\Semiring])$ equal to $1$, and all others equal to~$0$. Hence exactly one $\ms a_i$ is $[\1[\Semiring]]$, and all others are empty.
	
	\item To prove locally finite additive refinement, take locally finite families $(\ms a_i)_{i\in I}$, $(\ms b_j)_{j\in J}$ and labellings $\lambda:I\to X$, $\mu:J\to X$, such that
	\[
		\forall x\in X,\quad
		\mssum_{\substack{i\in I\\\lambda(i)=x}}\ms a_i
		=
		\mssum_{\substack{j\in J\\\mu(j)=x}}\ms b_j.
	\]
	For every $x\in X$ and $c\in\Semiring*$, this gives an equality in $\Nat$,
	\[
		\SUM_{\substack{i\in I\\\lambda(i)=x}}\ms a_i(c)
		=
		\SUM_{\substack{j\in J\\\mu(j)=x}}\ms b_j(c),
	\]
	with only finitely many nonzero summands. Apply $n$-ary additive refinement in $\Nat$ to each of these equalities, and put $\ms c_{ij}(c)=0$ when $\lambda(i)\neq\mu(j)$. We obtain maps $\ms c_{ij}:\Semiring*\to\Nat$ such that
	\begin{equation}
	\label{eq:the:msgrades-is-a-resource-semiring:additive-refinement}
		\forall i,c,\quad
		\ms a_i(c)=\SUM_j\ms c_{ij}(c),\qquad
		\forall j,c,\quad
		\ms b_j(c)=\SUM_i\ms c_{ij}(c).
	\end{equation}
	We must check that $(\ms c_{ij})_{i,j}$ is locally finite. If $\ms c_{ij}(c) \neq 0$, then both $\ms a_i(c) \neq 0$ and $\ms b_j(c) \neq 0$. Outside each neighbourhood $U$ of $\0[\Semiring]$, there are finitely many such pairs $(i,c)$ and $(j,c)$, hence finitely many triples $(i,j,c)$. By \cref{rem:currying}, the maps $\ms c_{ij}$ belong to $\lfmsgrades$ and form a locally finite family. The required identities follow by \cref{eq:the:msgrades-is-a-resource-semiring:additive-refinement}.
	
	\item For simultaneous locally finite mixed refinement, take locally finite families $(\ms a_k)_{k\in K}$, $(\ms b_{xk})_{x\in X}$ for every $k\in K$, and $(\ms c_i)_{i\in I}$, and a labelling $\lambda:I\to X$, such that
	\begin{equation}
	\label{eq:the:msgrades-is-a-resource-semiring:mixed-refinement-hypo}
		\forall x\in X,\quad
		\mssum_{k\in K}\ms a_k\mstimes\ms b_{xk}
		=
		\mssum_{\substack{i\in I\\\lambda(i)=x}}\ms c_i.
	\end{equation}
	We use the decomposition of natural coefficients into units to obtain a common decomposition of each $\ms a_k$ into singletons, independently of $x$. As in the proof of mixed refinement in $\Nat$, the decompositions of each $\ms b_{xk}$ are then obtained by $n$-ary additive refinement.
	
	Decompose each natural coefficient $\ms a_k(a)$ into a sum of $\ms a_k(a)$ units. Explicitly, put
	\[
		L_k\eqdef
		\set{(a,r)}
		[a\in\Semiring*,\ r\in\set{1,\dots,\ms a_k(a)}],
		\qquad
		a_{k,(a,r)}\eqdef a,\qquad
		\ms a_{kl}\eqdef[a_{kl}].
	\]
	Then $\ms a_k=\mssum_{l\in L_k}\ms a_{kl}$. By \cref{rem:currying}, the family $(a_{kl})_{k,l}$ is locally finite in $\Semiring$, hence $(\ms a_{kl})_{k,l}$ is locally finite in $\lfmsgrades$.
	
	Now, for fixed $x\in X$ and $c\in\Semiring*$, define
	\begin{align*}
		I_{xc} & \eqdef
		\set*{i\in I}[\lambda(i)=x,\ \ms c_i(c) \neq 0],
		\\
		J_{xc} &\eqdef
		\set*{(k,l,b)}
		[k\in K,\ l\in L_k,\ b\in\Semiring*,\
		a_{kl}\*[\Semiring]b=c,\ \ms b_{xk}(b) \neq 0].
	\end{align*}
	These sets are finite.
	Indeed, local finiteness of $(\ms c_i)_i$ gives finiteness of $I_{xc}$. 
	\thmitemref{def:lf-families-sums:product} gives local finiteness of $(\ms a_k\mstimes\ms b_{xk})_{k,x}$, hence there are finitely many $k,x$ such that $(\ms a_k\mstimes\ms b_{xk})(c) \neq 0$;
	each of these is a finite sum of summands $\ms a_k(a) \times \ms b_{xk}(b)$, for some $a \*[\Semiring] b = c$; 
	each such $a$ is equal to finitely many $a_{kl}$.
	This gives finiteness of $J_{xc}$.
	
	\Cref{eq:the:msgrades-is-a-resource-semiring:mixed-refinement-hypo}, evaluated at $c$, becomes
	\[
		\SUM_{(k,l,b)\in J_{xc}}\ms b_{xk}(b)
		=
		\SUM_{i\in I_{xc}}\ms c_i(c).
	\]
	Apply $n$-ary additive refinement in $\Nat$ to this equality, for every $x,c$. This defines a coefficient $\ms b_{ikl}(b)$ whenever $x=\lambda(i)$ and $c=a_{kl}\*[\Semiring]b$; put $\ms b_{ikl}(b) \eqdef 0$ otherwise. They satisfy
	\begin{gather*}
		\forall x,k,l,b,\quad
		\ms b_{xk}(b)
		=
		\SUM_{\substack{i\in I\\\lambda(i)=x}}\ms b_{ikl}(b),\qquad
		\forall i,c,\quad
		\ms c_i(c)
		=
		\SUM_{\substack{k\in K,\ l\in L_k,\ b\in\Semiring*\\
			a_{kl}\*[\Semiring]b=c}}
		\ms b_{ikl}(b).
	\end{gather*}
	
	We must now prove that $(\ms b_{ikl})_{i\in I}$ is locally finite for every $k,l$. Fix $k\in K$, $l\in L_k$ and a neighbourhood $U$ of $\0[\Semiring]$. By \cref{rem:currying}, local finiteness of $(\ms b_{xk})_x$ gives finitely many pairs $(x,b)$ with $b\notin U$ and $\ms b_{xk}(b)\neq0$.
	For any $\ms b_{ikl}(b)\neq0$, we must have $\ms b_{\lambda(i)k}(b)\neq0$;
	for each of the finitely many pairs $(x,b)$ above, a nonzero $\ms b_{ikl}(b)$ with $\lambda(i)=x$ requires, by construction, $i\in I_{x,a_{kl}\*[\Semiring]b}$, a finite set. Consequently
	\[
		F_U\eqdef
		\set{(i,b)\in I\times(\Semiring*\setminus U)}
		[\ms b_{ikl}(b)\neq0]
	\]
	is finite.
	For each $i$, this implies that $\support(\ms b_{ikl})\setminus U$ is finite. Since $U$ was arbitrary, each $\ms b_{ikl}$ belongs to $\lfmsgrades$. Moreover, $\set{i\in I}[\ms b_{ikl}\notin V_U([])]$ is equal to the first projection of $F_U$, hence is finite. Thus $(\ms b_{ikl})_{i\in I}$ is locally finite.
	
	In addition, we obtain:
	\[
		\forall x,k,l,\quad
		\ms b_{xk}
		=
		\mssum_{\substack{i\in I\\\lambda(i)=x}}\ms b_{ikl},
		\qquad
		\forall i,\quad
		\ms c_i
		=
		\mssum_{\substack{k\in K\\l\in L_k}}
		\ms a_{kl}\mstimes\ms b_{ikl}.
	\]
	\end{itemize}
	
	Finally, let $G$ be a guarding set of $\Semiring$, with associated continuous morphism $\gamma:\Semiring\to\Bool$. Define
	\[
		\msgset\gamma(\ms a)
		\eqdef
		\bigvee_{a\in\support(\ms a)}\gamma(a).
	\]
	Only finitely many terms are nonzero, since $G$ is a neighbourhood of $\0[\Semiring]$. Positivity and absence of zero divisors in $\Nat$, together with the sum and product equations for $\gamma$, give
	$\msgset\gamma(\ms a\msplus\ms b)
	= \msgset\gamma(\ms a)\vee\msgset\gamma(\ms b)$ and
	$\msgset\gamma(\ms a\mstimes\ms b)
	= \msgset\gamma(\ms a)\wedge\msgset\gamma(\ms b)$.
	Moreover $\msgset\gamma([])=0$ and $\msgset\gamma([\1[\Semiring]])=1$. Thus $\msgset\gamma$ is a unital semiring morphism with $\msgset\gamma^{-1}(\set{0})=\msgset G$. It is constant on every $V_G(\ms a)$, hence continuous. Therefore $\msgset G$ is a guarding set of $\lfmsgrades$.
	\end{proof}

Consequently, subject reduction and expansion hold in $\lfitypes[\lfmsgrades]$ and $\lfritypes[\lfmsgrades]$, by \cref{the:srse-locally-finite,the:lf-srse-relevant}.

\begin{remark}
	One could replace $\Nat$ as a target of multisets by another semiring $\Quantities$, using its operations for multiplicities. The proof of \cref{the:msgrades-is-a-resource-semiring} then requires appropriate refinement and discreteness properties of $\Quantities$. In particular for $\Quantities=\Bool$, the construction only produces a \emph{weak} locally finite resource semiring.
\end{remark}

\begin{theorem}[Collapse]
	\label{the:collapse}
	The \emph{collapse} surjective semiring morphism defined and extended to (multi-)types as in \cref{the:finite-collapse} translates typings in $\lfitypes[\lfmsgrades]$ into typings in $\lfitypes[\Semiring]$, and likewise for the relevant systems. If $\Semiring$ has positivity, this translation preserves positive typings.
\end{theorem}

	\begin{proof}
	The family $(a_l)_l$ is locally finite, so its sum exists and $\kappa$ is well-defined. Preservation of addition and multiplication are immediate by the locally finite semiring laws of~$\Semiring$. 
	As for preservation of locally finite sums,
	let $(\ms a_i)_{i\in I}$ be locally finite, with $\ms a_i=[a_{il}]_{l\in L_i}$. By \cref{rem:currying}, $(a_{il})_{i,l}$ is locally finite. By \thmitemref{def:lf-families-sums:partition}, $(\collapse(\ms a_i))_{i\in I}$ is locally finite and
	$\collapse\left(\mssum_{i\in I}\ms a_i\right)
	=\SUM_{i,l}a_{il}
	=\SUM_{i\in I}\collapse(\ms a_i)$.
	
	Continuity of $\gamma$ gives $\gamma(\collapse(\ms a))=\msgset\gamma(\ms a)$, hence $\collapse^{-1}(G)=\msgset G$. As a consequence, an immediate induction using all these properties concludes that valid typings in $\lfitypes[\lfmsgrades]$ are translated to valid typings in $\lfitypes$.
	Preservation of positive typings follows from the same argument as in \cref{the:finite-collapse}.
	\end{proof}

\begin{remark}
	In particular, $\fmsgrades$ is a subsemiring of $\lfmsgrades$. If $\0[\Semiring]$ is isolated in $\Semiring$, the condition of \cref{def:lfmsgrades} applied to the neighbourhood $\set{\0[\Semiring]}$ forces every multiset of grades to have finite support, so $\fmsgrades=\lfmsgrades$. Moreover, $V_{\set{\0[\Semiring]}}(\ms a)=\set{\ms a}$ for every $\ms a\in\lfmsgrades$, so the topology defined on $\lfmsgrades$ is discrete.
\end{remark}

\begin{example}
	\label{exa:multiset-constr-tropical}
	Let us give some details in the case $\Semiring=\Tropical$:
	
	\begin{itemize}
	\item The monoid $\Tropical*$ is the usual $(\natset,+,0)$.
	
	\item $\Tropicalhat$ is just $\Tropical*\to\Nat$. Indeed, a neighbourhood basis of $\infty$ in $\Tropical$ is given by $U_d\eqdef\set{\infty}\cup\set{c\in\Tropical*}[c>d]$, for $d \in \natset$. All these sets have finite complements.
	
	\item The operations on $\Tropicalhat$ are
	\[
		(\ms a\msplus\ms b)(c)\eqdef\ms a(c)+\ms b(c),\qquad
		(\ms a\mstimes\ms b)(c)\eqdef\SUM_{a+b=c}\ms a(a)\ms b(b).
	\]
	The zero and unit are $[]$ and $[0]$, respectively.
	
	\item The topology of $\Tropicalhat$ introduced in \cref{def:lfmsgrades} is the one induced by the ultrametric
	\[
		\dist[\Tropicalhat](\ms a,\ms b)\eqdef
		\begin{cases}
			0&\text{if }\ms a=\ms b,\\
			2^{-\min\set{c}[\ms a(c)\neq\ms b(c)]}&\text{otherwise.}
		\end{cases}
	\]
	Indeed, $V_{U_d}(\ms a)$ consists of the multisets agreeing with $\ms a$ at grades at most $d$.
	
	\item The guarding set introduced in \cref{the:msgrades-is-a-resource-semiring} is $G_{\Tropicalhat} \eqdef \msgset{G_{\Tropical}} = \set{\ms a}[\ms a(0)=0]$.
	
	\item By \cref{rem:currying}, the local finiteness condition on multi-types $\ms m : \lfitypes[\Tropicalhat] \to \Tropicalhat$ is
	\[
		\forall d\in\natset,\ 
		\set*{(\sigma,c)\in \lfitypes[\Tropicalhat]\times\Tropical*}[c\<= d,\ \mscurr m(\sigma,c)\neq0]\text{ is finite},
	\]
	\ie, within any delay $d$, $\ms m$ contains only finitely many types.
	
	\item The collapse map from \cref{the:collapse} is given by $\collapse(\ms a) = \min(\support(\ms a))$.
	\end{itemize}
\end{example}

\begin{remark}[Finite and Locally Finite Mixed Refinement]
	\label{rem:finite-and-locally-finite-refinement}
	A locally finite resource semiring need not be a resource semiring: the refining family in mixed refinement may have to be infinite, even when the hypothesis is a single binary equality. This happens in $\Tropicalhat$, as follows.
	
	Put $a_p=4^p$, $b_p=2\times 4^p$ for $p\in\natset$, and
	\[
		\ms a=[a_p]_p,\qquad \ms b=[b_q]_q,\qquad
		\ms c_1=[a_p+b_p]_p,\qquad \ms c_2=[a_p+b_q]_{p\neq q}.
	\]
	These multisets belong to $\Tropicalhat$, and $\ms a\mstimes\ms b=\ms c_1\msplus\ms c_2$. All grades $a_p+b_q$ are distinct: for $p\neq q$ the base-$4$ expansion of $a_p+b_q$ has digit $1$ in position $p$ and digit $2$ in position $q$; for $p=q$, it has digit $3$ in position $p$. All other digits are zero. Thus $a_p+b_q$ uniquely determines $(p,q)$.
	
	Now, suppose that mixed refinement gave a finite set $L$ and multisets satisfying
	\[
		\ms a=\mssum_{l\in L}\ms a_l,\qquad
		\forall l\in L,\ \ms b=\ms b_{1l}\msplus\ms b_{2l},\qquad
		\forall i\in\set{1,2},\ \ms c_i=\mssum_{l\in L}\ms a_l\mstimes\ms b_{il}.
	\]
	Observe that for all $i$, $l$, $p$ and $q$,
	\begin{equation*}
		(\ms a_l \mstimes \ms b_{il})(a_p + b_q)
		= \sum_{\substack{a,b \\ a+b=a_p+b_q}} \ms a_l(a) \ms b_{il}(b) \\
		= \sum_{\substack{p',q' \\ a_{p'}+b_{q'}=a_p+b_q}} \ms a_l(a_{p'}) \ms b_{il}(b_{q'})
		= \ms a_l(a_p) \ms b_{il}(b_q)
	\end{equation*}
	because $\support(\ms a_l) \subseteq \support(\ms a)$
	and $\support(\ms b_{il}) \subseteq \support(\ms b)$,
	and because $a_p+b_q$ uniquely determines $(p,q)$.
	
	Since $\ms a(a_p)=1$, there is exactly one $l(p)\in L$ such that $\ms a_{l(p)}(a_p)=1$, and $\ms a_l(a_p) = 0$ for any other $l$.
	Hence for all $p$ and $q$,
	\begin{equation*}
		\ms c_1(a_p+b_q) = \sum_{l \in L} \ms a_l(a_p) \ms b_{1l}(b_q)
		= \ms b_{1,l(p)}(b_q).
	\end{equation*}
	
	Suppose there are $p \neq q$ such that $l(p) = l(q)$, then
	\begin{equation*}
		1 = \ms c_1(a_p+b_p) = \ms b_{1,l(p)}(b_p)
		= \ms b_{1,l(q)}(b_p) = \ms c_1(a_q+b_p) = 0.
	\end{equation*}
	Thus $p\mapsto l(p)$ is injective, which is impossible for finite $L$.
	(Instead, a locally finite refinement does exist: take $L=\natset$, $\ms a_p=[a_p]$, $\ms b_{1p}=[b_p]$ and $\ms b_{2p}=[b_q]_{q\neq p}$. The family $(\ms a_p)_p$ is locally finite, and the displayed identities hold.)
	
	Consequently, subject reduction (and expansion) of the system $\lfitypes[\Tropicalhat]$ does hold by \cref{the:srse-locally-finite}, whereas subject reduction of the finite system $\itypes[\Tropicalhat]$ does not: there is derivation
	\begin{multline*}
		h : [\gr {[\gr{α}{\ms a}] \lto α} {[0]}],
		g : [\gr {[\gr{α}{\ms b}] \lto α} {\ms a}],
		f : \left[\begin{array}{l}
			\gr {[] \lto α} {\ms c_1}, \\
			\gr {[\gr{α}{[0]}] \lto α} {\ms c_2}
			\end{array}\right],
		x : [\gr{α}{\ms c_2}] \\
		\vdash (λy.h(gy))(fx) : α,
	\end{multline*}
	but no finite typing of $h(g(fx))$ with the same type and context: the typing of the redex crucially relies on the fact that the abstracted variable $y$ has multi-types $\ms a\mstimes\ms b$ \emph{and} $\ms c_1\msplus\ms c_2$.
\end{remark}

\section{Operational Characterisations}
\label{sec:operational}

One main interest of usual intersection types is that they do not only guarantee, but also \emph{characterise} some operational properties of λ-terms. Classically, typing in $\itypes[\Bool]$ and $\itypes[\Nat]$ characterises head normalisation \autocite{Salle.80,Salle.80a,deCarvalho.07}: we show that this can actually be extended to all systems defined above. We then recall two other characterisations from the literature, to illustrate how restricting the shape of the types allows to account for other notions of termination.

\subsection{Head Normalisation}

Write $\HN$ for the terms having a head normal form $\lambda x_1\dots x_n.yM_1\dots M_k$.

\begin{definition}
	A \emph{reducibility candidate} is a set $\rcstyle{R}\subseteq\HN$ 
	containing all terms $xM_1\dots M_k$ and 
	closed under inverse head contraction: if $P\subst Q\,\vec M\in \rcstyle{R}$ then $(\lambda x.P)Q\,\vec M\in \rcstyle{R}$.
\end{definition}

 The set $\HN$ itself is a candidate. Any nonempty intersection of candidates is a candidate. For any set $\Var\subseteq R\subseteq\lamterms$ and candidate $\rcstyle{S}$,
\[
	R\to \rcstyle{S}\eqdef\set{M}[\forall N\in R,\ MN\in \rcstyle{S}]
\]
is a candidate too (this holds in particular when $R$ is a candidate).

\begin{theorem}[Soundness and Completeness \wrt Head Normalisation]
	\label{the:hn}
	For any positive semiring $\Semiring$ with no zero divisors, $\itypes$ and $\ritypes$ characterise head normalisation: $M$ is typable \ifandonlyif $M\in\HN$.
	The same holds for $\lfitypes$ and $\lfritypes$ if $\Semiring$ admits locally finite sums and has a guarding set $G$.
\end{theorem}

	\begin{proof}
	\emph{Soundness.}
	For the full systems, let $G$ be the guarding set defined on $\Semiring$, and $\gamma : \Semiring \to \Bool$ the associated morphism (as in \cref{def:guarding-set}). For the finite systems, use the morphism $\gamma : \0 \mapsto 0$, and $a \mapsto 1$ for $a \neq \0$; positivity and absence of zero divisors make it a semiring morphism.
	Given an environment $\rho$ mapping atoms to candidates, define the reducibility interpretation of a type or a multi-type by:
	\[
		\redinterp\alpha\eqdef\rho(\alpha),\qquad
		\redinterp{\ms m\lto\sigma}\eqdef
		\redinterp{\ms m}\to\redinterp\sigma,\qquad
		\redinterp{\ms m}\eqdef
		\bigcap_{\gamma(\ms m(\sigma))=1}\redinterp\sigma.
	\]
	For finite types this definition is well-founded by induction on their formation trees. For the full systems it is well-founded since we recurse only in inductive positions: the condition $\gamma(\ms m(\sigma))=1$ exactly means that $\sigma$ is not guarded by its coefficient in $\ms m$.
	(An empty intersection is defined to be $\lamterms$.)
	The semiring equations, and for the full systems the continuity of $\gamma$ (ensuring that it preserves locally finite sums), give:
	\[
		\redinterp{\ms m\+\ms n}
		=\redinterp{\ms m}\cap\redinterp{\ms n},\qquad
		\redinterp*{\SUM_{i\in I} a_i\*\ms m_i}
		=\bigcap_{\substack{i \in I \\\gamma(a_i)=1}}\redinterp{\ms m_i}.
	\]
	Using these identities, a straightforward induction on the derivation (in either system) allows to prove that
		for all derivation $\derives x_1:\ms m_1,\dots,x_n:\ms m_n\vdash M:\sigma$, 
		for all $N_i\in\redinterp{\ms m_i}$, for $1\<= i\<= n$,
		one has $M\subst[\vec x]{\vec N}\in\redinterp\sigma$,
	and likewise for multi-types. 
	Finally, take $\rho(\alpha) \eqdef \HN$ and $N_i \eqdef x_i$ to obtain $M\in\redinterp\sigma\subseteq\HN$ for any typable $M$.

	\emph{Completeness.} Type a head normal form $\lambda\vec x.yM_1\dots M_k$ by assigning $y$ the type $[]\lto\dots\lto[]\lto\alpha$, with $k$ empty domains, typing each argument with $[]$, then applying \drule{\lto_e} and \drule{\lto_i}. Conclude by subject expansion in the finite system (which holds without any assumption on $\Semiring$, as a consequence of \cref{lem:backward-substitution}). The resulting derivation is also valid in the full system.
\end{proof}

\begin{remark}
\label{rem:hn-soundness-in-I2-completeness-in-rIN}
	Actually it enough to prove soundness in $\itypes[\Bool]$ and completeness in $\ritypes[\Nat]$:
	\begin{itemize}
	\item for soundness, any derivation $\derives \Gamma \vdash M:\sigma$ in $\itypes$ (or $\lfitypes$) can be turned into a derivation in $\itypes[\Bool]$ by applying $\gamma : \Semiring \to \Bool$ to all coefficients appearing in the derivation;
	\item for completeness, any head normalising term can be typed in $\ritypes[\Nat]$, by typing its head normal form as above and applying subject expansion, and the resulting derivation can be turned into a derivation in $\ritypes[\Semiring]$ by applying to all its coefficients the semiring morphism $\Nat \to \Semiring$, $n \mapsto \1 \+ \dots \+ \1$ ($n$ times).
	\end{itemize}
\end{remark}

\subsection{Positive Typing}

Let us recall two further characterisations from the literature, which use restrictions to \emph{positive} typings.

\begin{definition}[Positive Typing]
	The polarity of a position in the syntax tree of a type is defined as follows:
	\begin{itemize}
	\item the root of the type is positive,
	\item polarity is reversed when entering the domain of an arrow $\lto$,
	\item polarity is preserved when entering the codomain of an arrow, and when entering in an element of an $\Semiring$-multiset.
	\end{itemize}
	A typing derivation $\derives \Gamma \vdash M:\sigma$ is \emph{positive} if $[]$ does not occur in any $\Gamma(x)$ in negative position, and in $\sigma$ in positive position.
\end{definition}

Write $\WN$ for the set of all λ-terms having a β-normal form.
The following Theorem extends the well-established characterisation of weak β-normalisation by positive typing in $\itypes[\Bool]$ \autocite{Coppo.Dez.78,Krivine.90} and in $\ritypes[\Nat]$ \autocite{deCarvalho.07}.

\begin{theorem}[Soundness and Completeness \wrt Weak β-Normalisation]
	\label{the:wn}
	For any positive semiring $\Semiring$ with no zero divisors, positivie typing in $\itypes$ and $\ritypes$ characterise weak normalisation: $M$ is typable \ifandonlyif $M\in\WN$.
\end{theorem}

	\begin{proof}
	As in \cref{rem:hn-soundness-in-I2-completeness-in-rIN}, we can use soundness of $\itypes[\Bool]$ and completeness of $\ritypes[\Nat]$:
	\begin{itemize}
	\item for soundness, instead of using $\gamma$, apply to all coefficients the semiring morphism $\chi:\Semiring\to\Bool$ defined by $\chi(a)=0$ \ifandonlyif $a=\0$. The resulting derivation is a positive derivation in $\itypes[\Bool]$: since $\chi$ preserves zero and nonzero coefficients (this is a consequence of positivity and absence of zero divisors), the translation preserves empty and nonempty multi-types;
	\item for completeness, any $M\in\WN$ has a positive derivation in $\ritypes[\Nat]$. Apply to all its coefficients the semiring morphism $\Nat\to\Semiring$, $n\mapsto\1\+\dots\+\1$ ($n$ times). Positivity ensures that nonzero coefficients remain nonzero, so the resulting derivation in $\ritypes[\Semiring]$ is positive.
	\end{itemize}
	The restriction to the finite systems is essential, since $\chi$ need not preserve local finiteness of multisets or guardedness of types.
	\end{proof}

Write $\HHN$ for the set of all hereditary head normalising λ-terms, \ie $\HHN$ is the coinductively defined set such that $M \in \HHN$ \ifandonlyif $M \in \HN$ and, if $M \bred* \lambda x_1\dots x_n.yM_1\dots M_k$, then $M_i \in \HHN$ for all $1 \le i \le k$. Equivalently, $M \in \HHN$ is its Böhm tree contains no undefined subterm $\bot$. The following characterisation was given in \autocite{Cerda.Dal.26}.

\begin{theorem}[Soundness and Completeness \wrt Hereditary Head Normalisation]
	\label{the:hhn}
	Positive typing in $\lfitypes[\Tropical]$ characterises $\HHN$: $M$ is typable by a positive derivation \ifandonlyif $M \in \HHN$.
\end{theorem}

The lifted system $\lfritypes[\Tropicalhat]$ from \cref{exa:multiset-constr-tropical} is exactly the \enquote{auxiliary system} used in \autocite{Cerda.Dal.26} to prove \cref{the:hhn}: one first proves completeness of positive typings in $\lfritypes[\Tropicalhat]$ \wrt hereditary head normalisation, which immediately gives completeness of positive typings in $\lfitypes[\Tropicalhat]$ by inclusion, and finally in $\lfitypes[\Tropical]$  by collapse (\cref{the:collapse}).

\section*{Acknowledgements}

Tito Nguyễn pointed us at an error in a first version of this draft and drew our attention to parts of the literature around \cref{def:resource-semiring}, which sparked insightful discussions with Quentin Aristote and Vincent Moreau. He also suggested the counter-example λ-term from \cref{rem:finite-and-locally-finite-refinement}, which has a semantic counterpart in \autocite{Moreau.Ngu.26}. Fruitful exchanges also took place with Gabriele Vanoni. The authors wish to thank them all.

\printbibliography

\end{document}